\documentclass[a4paper,11pt]{article}

\usepackage{jinstpub}

\usepackage{lineno}

\usepackage{comment}
\usepackage[version=4]{mhchem}
\usepackage{siunitx}
\usepackage{rotating}
\usepackage{tabularx}
\usepackage{ctable}
\usepackage{hyperref}
\usepackage{doi}
\usepackage{afterpage}

\usepackage{multicol}

\renewcommand{\unit}[1]{\ensuremath{\;\mathrm{#1}}}
\newcommand{\software}[1]{\texttt{#1}}
\newcommand{\CRY}{\software{CRY}}
\newcommand{\geant}{\software{GEANT4}}
\newcommand{\iso}[2]{\ensuremath{\ce{^{#2}#1}}}
\newcommand{\dims}[3]{#1$\times$#2$\times$#3}

\title{Abatement of Ionizing Radiation for Superconducting Quantum Devices}

\author[a,1]{B.~Loer,\note{Corresponding author.}} 
\author[b]{P.~M.~Harrington,} 
\author[a]{B.~Archambault,} 
\author[a]{E.~Fuller,} 
\author[a]{B.~Pierson,} 
\author[a]{I.~Arnquist,} 
\author[a]{K.~Harouaka,} 
\author[a]{T.~D.~Schlieder,} 
\author[c]{D.~K.~Kim,} 
\author[c]{A.~J.~Melville,} 
\author[c]{B.~M.~Niedzielski,} 
\author[c]{J.~L.~Yoder,} 
\author[b,c]{K.~Serniak,} 
\author[b,c]{W.~D.~Oliver,} 
\author[a]{J.~L.~Orrell,} 
\author[a]{R.~Bunker,}
\author[a]{B.~A.~VanDevender,} 
\author[a]{M.~Warner} 

\affiliation[a]{Pacific Northwest National Laboratory,\\ 902 Battelle Boulevard, Richland, WA, USA}
\affiliation[b]{Research Laboratory of Electronics, Massachusetts Institute of Technology, Cambridge, MA 02139, USA}
\affiliation[c]{MIT Lincoln Laboratory,\\ 244 Wood Street, Lexington, MA, USA}

\emailAdd{ben.loer@pnnl.gov}

\abstract{
Ionizing radiation has been shown to reduce the performance of superconducting quantum circuits. In this report, we evaluate the expected contributions of different sources of ambient radioactivity for typical superconducting qubit experiment platforms.
Our assessment of radioactivity inside a typical cryostat highlights the importance of selecting appropriate materials for the experiment components nearest to qubit devices, such as packaging and electrical interconnects. We present a shallow underground facility (30-meter water equivalent) to reduce the flux of cosmic rays and a lead shielded cryostat to abate the naturally occurring radiogenic gamma-ray flux in the laboratory environment. We predict that superconducting qubit devices operated in this facility could experience a reduced rate of correlated multi-qubit errors by a factor of approximately 20 relative to the rate in a typical above-ground, unshielded facility. 
Finally, we outline overall design improvements that would be required to further reduce the residual ionizing radiation rate, down to the limit of current generation direct detection dark matter experiments.

}


\begin{document}
\setcounter{tocdepth}{2}
\maketitle
\flushbottom



\section{Introduction}\label{sec:intro}

Quantum technologies that leverage entanglement between multiple sensors or computing elements (qubits) have the potential to dramatically advance a range of computing and sensing applications~\cite{preskill2018, douglasbeck2019, cloet2019, carlson2018, degen2017a}. Many different technologies are being investigated for the physical implementation of qubits, but much focus has been placed on superconducting qubits due, in part, to their ease of manufacturing with standard semiconductor fabrication techniques as well as control and readout with microwave pulses~\cite{hassler2011, oliver2013, kjaergaard2020}. A key characteristic affecting the real-world computing potential of qubits of any modality is the coherence time---how long on average a qubit will remain in a given quantum state. Improving the coherence time of superconducting qubits has been a major research focus for the past several years~\cite{kjaergaard2020}. 

Recent experiments have demonstrated that ionizing radiation can directly lead to superconducting qubit decoherence~\cite{vepsalainen2020}. 
Notably, error ``bursts'' that are correlated in time across multiple qubits and extending over entire device substrates have been observed with characteristics consistent with the production of nonequilibrium quasiparticles by ionizing radiation~\cite{cardani2021, chen2021, wilen2021, mcewen2022, acharya2023, thorbeck2023}. Correlated error events have been shown to occur in part from cosmic-ray impacts~\cite{harrington2024, li2024}.
This poses a challenge for the implementation of many proposed quantum error correction techniques, such as the surface code, which rely on an assumption of random and uncorrelated errors in space and time~\cite{shor1995, steane1996, klesse2005, fowler2012}.

In this report, we present an estimate of the rate of energy injections from sources of ionizing radiation into a typical device operating inside a dilution refrigerator, followed by a specific approach to reducing that rate. Ionizing radiation sources are separated into three components based on effective techniques for mitigation: (1) cosmic-ray-induced radiation, which can be reduced by operating in an underground location; (2) terrestrial gamma rays in the laboratory environment, which can be mitigated by surrounding the dilution refrigerator with a lead radiation shield; and (3) naturally-occurring radioactive isotopes in materials inside the dilution refrigerator, which can be abated by replacing with more radiopure alternatives and by an internal gamma shield. 

We address the first two sources of radiation with a design for a Low Background Cryogenic Facility (LBCF), a radiation-shielded dilution refrigerator sited in a shallow underground laboratory at Pacific Northwest National Laboratory (PNNL), which will allow operation of superconducting devices, such as qubits, with reduced ionizing radiation exposure. This concept represents one mitigation strategy among many that could work in concert with device design improvements (``radiation-hardening'') to realize superconducting qubits that are less impacted by the effects of ionizing radiation. 
This facility provides a near-term opportunity for research into ionizing radiation effects within a controlled environment. In particular we highlight the potential to quantify the effects of radiation on superconducting qubit performance, such as the average decoherence rate~\cite{vepsalainen2020}, the rate of spatiotemporally correlated qubit errors~\cite{wilen2021, mcewen2022,harrington2024}, reconfiguration of two-level systems (TLS)~\cite{thorbeck2023}, and performance of error correction codes~\cite{chen2021, acharya2023}. The facility will also enable testing the efficacy of design modifications intended to mitigate the impact of ionizing radiation on device performance, such as the use of normal metals for phonon downconversion~\cite{riwar2016, iaia2022} or detecting and ``vetoing'' likely error states using classical sensors~\cite{orrell2021} or spectator qubits not directly participating in the quantum computation~\cite{acharya2023}.

\section{Estimating the radiation environment}
\subsection{Monte Carlo simulation}
\label{sec:simulationsetup}
We employ a Monte Carlo radiation-transport simulation to estimate how radiation fluxes measured in the laboratory interacts with devices inside a dilution refrigerator, with the goal of creating a radiation budget for a typical cryogenic device. The simulation uses the \geant{}~\cite{agostinelli2003, allison2006, allison2016} toolkit, version 10.7.p03. We simulate a generic laboratory space as a \dims{8}{8}{4}\unit{m^3} box. In this space various models of radiation-sensitive devices are simulated: a NaI detector used to measure the cosmic-ray muon flux (Section~\ref{sec:cosmicraysource}), a high purity germanium (HPGe) detector used to measure the environmental gamma flux (Section~\ref{sec:envgammasource}), and an array of silicon chips representing the substrates of superconducting quantum circuits in a simplified model of a dilution refrigerator. 

\begin{figure}
    \centering
    \includegraphics[width=\textwidth]{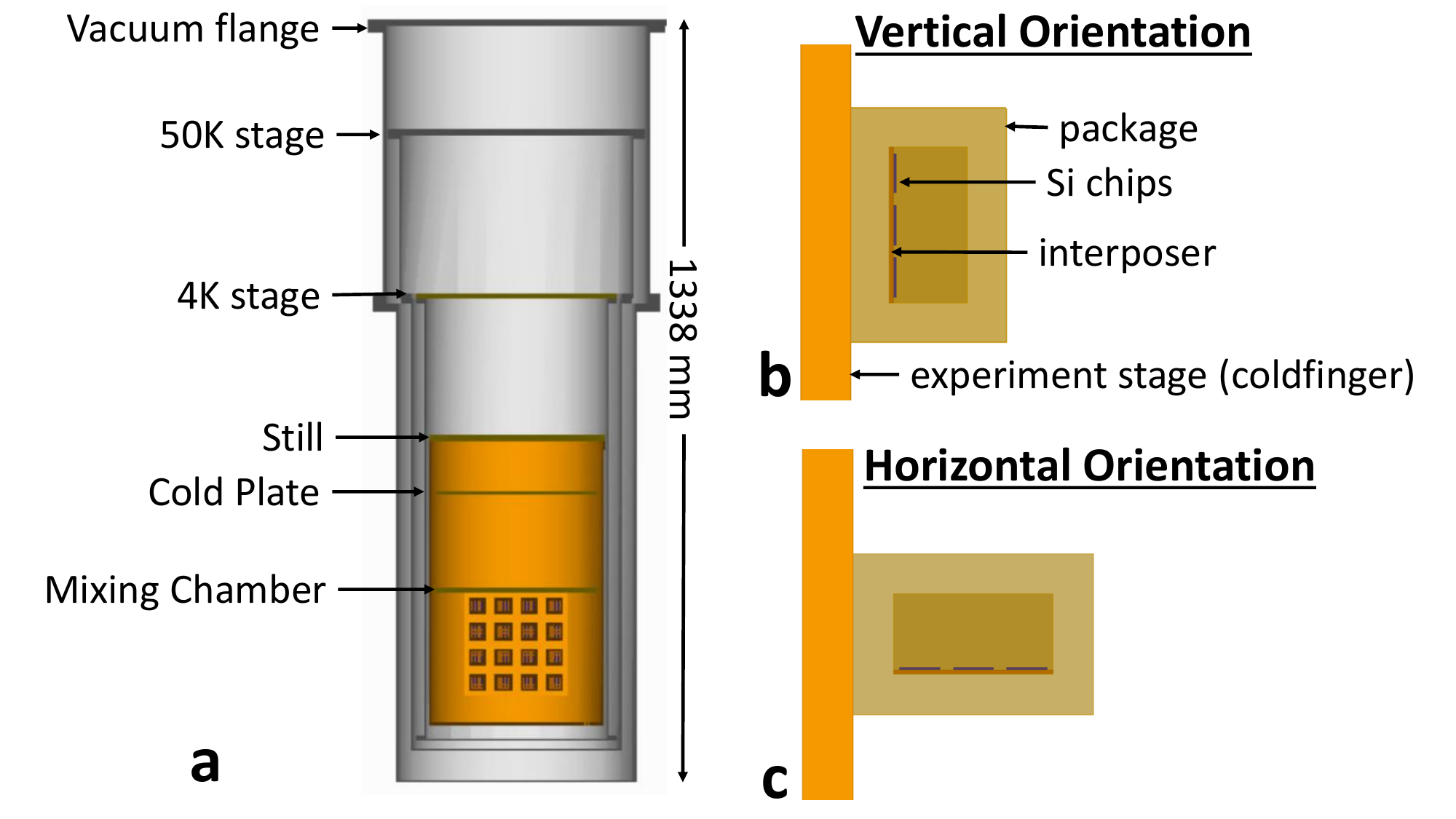}
    \caption{Cross-sectional view of the dilution refrigerator simulated in \geant{} showing the 16 device packages attached to a coldfinger experiment stage, and detailed views of the package in the (b) vertical and  (c) horizontal orientations.}
    \label{fig:simcomputer}
\end{figure}

We use a \dims{2.5}{5}{0.38}\unit{mm^3} silicon chip as a qubit substrate. We simulate the effect of radiation on 144 identical silicon chips within the cryostat to increase the computational efficiency of the radiation transport simulation, given the low probability of radiation interactions with any given chip. 
Groups of nine chips are placed on a 0.5~mm polyimide\footnote{Although polyimide is not typically used for this purpose for microwave frequency superconducting devices, it was convenient to implement because it is common in simulations of lower-frequency devices such as transition edge sensors. Due to the small mass, the details of this material choice will have negligible impact on the radiation transport.} ``interposer'' inside a \dims{3}{3}{2}\unit{cm^3} copper box with 0.5\unit{cm} thick sides that would act as an RF shielding package for typical microwave-addressed qubits. Sixteen of these units are attached to a 6.35~mm thick copper plate, 18~cm per side, hanging vertically from the dilution refrigerator mixing chamber stage. A \SI{1.59}{cm} thick cylindrical aluminum cylinder with inner diameter \SI{180.1}{mm} (just large enough to contain the plate) surrounds this plate (not shown in Figure~\ref{fig:simcomputer}). This cylinder is a placeholder for additional copper, aluminum, and mumetal shielding often included in superconducting qubit setups. The simulated dilution refrigerator is based approximately on the dimensions of a Bluefors LD-400 system, the dimensions of which are given in Table~\ref{tab:fridgedimensions}. 
Because cosmic-ray secondaries (primarily muons but also neutrons, protons, and gammas) are highly directional downward, the orientation of the silicon chips may have a significant effect on the interaction rate. Therefore when evaluating cosmic-ray effects, we perform separate simulations with the packages oriented either vertically (i.e., with the normal to the chip face along the horizon) or horizontally (with the normal vector pointed up toward the zenith). Figure~\ref{fig:simcomputer} shows the simulated dilution refrigerator and array of chips. 

\begin{table}[ht!]
    \centering
    \caption{Dimensions of stages and cans for the simulated dilution refrigerator. All dimensions are in millimeters. All copper materials are assumed to be gold-plated. For simplicity in the simulation, gold-plated copper is modeled as copper with 0.1\% bulk admixture of gold.}
    \label{tab:fridgedimensions}
    \begin{tabular}{|l|ccccl|}
    \multicolumn{6}{l}{\vspace{-0.75em}} \\
    \multicolumn{6}{l}{\textbf{Cooling stages}} \\ 
    \hline
    Stage & Vertical Offset & Radius & Thickness & & Material \\ 
    \hline
    Vacuum Flange & 0 & 261 & 12 & & stainless steel \\
    50K           & 191 & 223.5 & 12 & & aluminum \\
    4K            & 480 & 176 & 10 & & copper \\
    Still         & 730 & 153 & 9 & & copper \\
    Cold Plate (CP)   & 829 & 140 & 6 & & copper \\
    Mixing Chamber (MXC) & 997 & 142.3 & 8 & & copper \\
    \hline
    \multicolumn{6}{l}{\vspace{-0.75em}} \\
    \multicolumn{6}{l}{\textbf{Cans}} \\
    \hline
    Can & Vertical Offset & Radius & Height & Thickness & Material \\
    \hline
    Vacuum top & 12 & 230 & 486 & 3.2 & aluminum \\
    Vacuum bottom & 498 & 207.65 & 840 & 3.2 & aluminum \\
    50K top & 203 & 204 & 286.5 & 1 & aluminum \\
    50K bottom & 489.5 & 182 & 793 & 1 & aluminum \\
    4K & 490 & 160 & 774 & 1.5 & aluminum \\
    Still & 739 & 151.5 & 500 & 0.5 & copper \\
    \hline
    \end{tabular}
\end{table}


For each radiation source, we calculate three quantities for the qubit chips: the rate of interactions depositing greater than \SI{3}{eV} into the silicon substrate, the rate of interactions depositing greater than \SI{1}{MeV}, and the total absorbed dose (interaction rate times average interaction energy with no threshold) from all interactions. Each of these measures is normalized per unit mass of substrate. The lower threshold for counting interactions roughly corresponds to the silicon bandgap, and the precise value has negligible effect on the results presented. We expect the minimum energy injection required to observe a response response to be highly device-dependent. The very low threshold we have adopted is supported by measurements with SuperCDMS detectors in which eV-scale energy deposits excite sensors spread over a square centimeter on a gram-scale device~\cite{amaral2020}\footnote{Note that SuperCDMS detectors have a substantially different aspect ratio ($\sim$1:3 vs.\ $\sim$1:10 or smaller for a typical qubit chip) and are intentionally thermally isolated. Both of these features may lead to a more uniform phonon population in the substrate and thus greater sensitivity to energy deposits.}, and by  simulations with the GEANT4 Condensed Matter Physics (G4CMP) toolkit~\cite{kelsey2023}. 

\subsection{Cosmic-ray secondaries}
\label{sec:cosmicraysource}
For both external sources of radiation (cosmic-ray secondaries and environmental gammas), we use the same basic approach to estimate interaction rates in superconducting device substrates. First, we measure the energy spectrum of interactions with a commercial radiation detector. Then we simulate the appropriate particle fluxes interacting with a model of that detector in our radiation transport code and compare with the measured data to obtain an overall normalization constant for the absolute particle flux. Finally, we use the estimated absolute particle flux to simulate the rate of interactions in the simulated silicon chips. A similar procedure was previously used to estimate the ionizing radiation environment for a superconducting qubit chip~\cite{vepsalainen2020}.

The flux of cosmic-ray secondary particles 
depends strongly on atmospheric depth (or elevation relative to sea level) and any overburden (including upper building floors), and weakly on latitude and solar cycle~\cite{grieder2001}. For this study, we employ the \CRY{} cosmic-ray shower generator software~\cite{crypackage1} to generate distributions of cosmic-ray secondary particles that are then propagated in our \geant{} model. To validate and normalize the simulations, we measured the spectrum of cosmic-ray muon interactions in two laboratories at PNNL: a surface laboratory and the shallow underground laboratory (SUL)~\cite{aalseth2012}.  Spectra were acquired with a \SI{3}{inch} (7.62~cm) diameter and height NaI(Tl) scintillator attached to a photomultiplier tube (PMT) read out by a Mirion Osprey integrated base and multichannel analyzer (MCA)~\cite{mirionosprey}. The PMT was operated with lower-than-nominal high voltage to bring the $\sim$\SI{40}{MeV} minimum ionizing muon peak into the range of the MCA. Data was acquired for 92 hours in the surface laboratory and 410 hours in the SUL. 

To compare \CRY{} simulation results to the measurements, we add a simplified model of the NaI detector to our simulation described in Section~\ref{sec:simulationsetup}. The simulated detector is a NaI cylinder with \SI{3}{inch} diameter and height enclosed in a \SI{0.5}{mm} thick aluminum oxide ``reflector'' and \SI{0.8}{mm} thick aluminum case; the PMT is not modeled. The simulated detector is placed in the center of the laboratory model volume, 1.5~m from the floor. cosmic-ray secondary particles generated by \CRY{} are propagated, and the energy deposited in the simulated NaI volume is recorded and normalized to counts per second using the live time reported by \CRY. For the surface measurement, no walls or ceiling are modeled in the simulation, and the \CRY{}-generated cosmic-ray secondaries are produced from a square plane \SI{20}{m} on a side, \SI{4}{m} above the floor (\SI{2.5}{m} above the detector), which accounts for $\sim$95\% of the total cosmic-ray muon flux assuming a $\cos^2\theta$ zenith angle dependence~\cite{grieder2001}. 

\begin{figure}[ht!]
    \centering
    \includegraphics[width=0.85\textwidth]{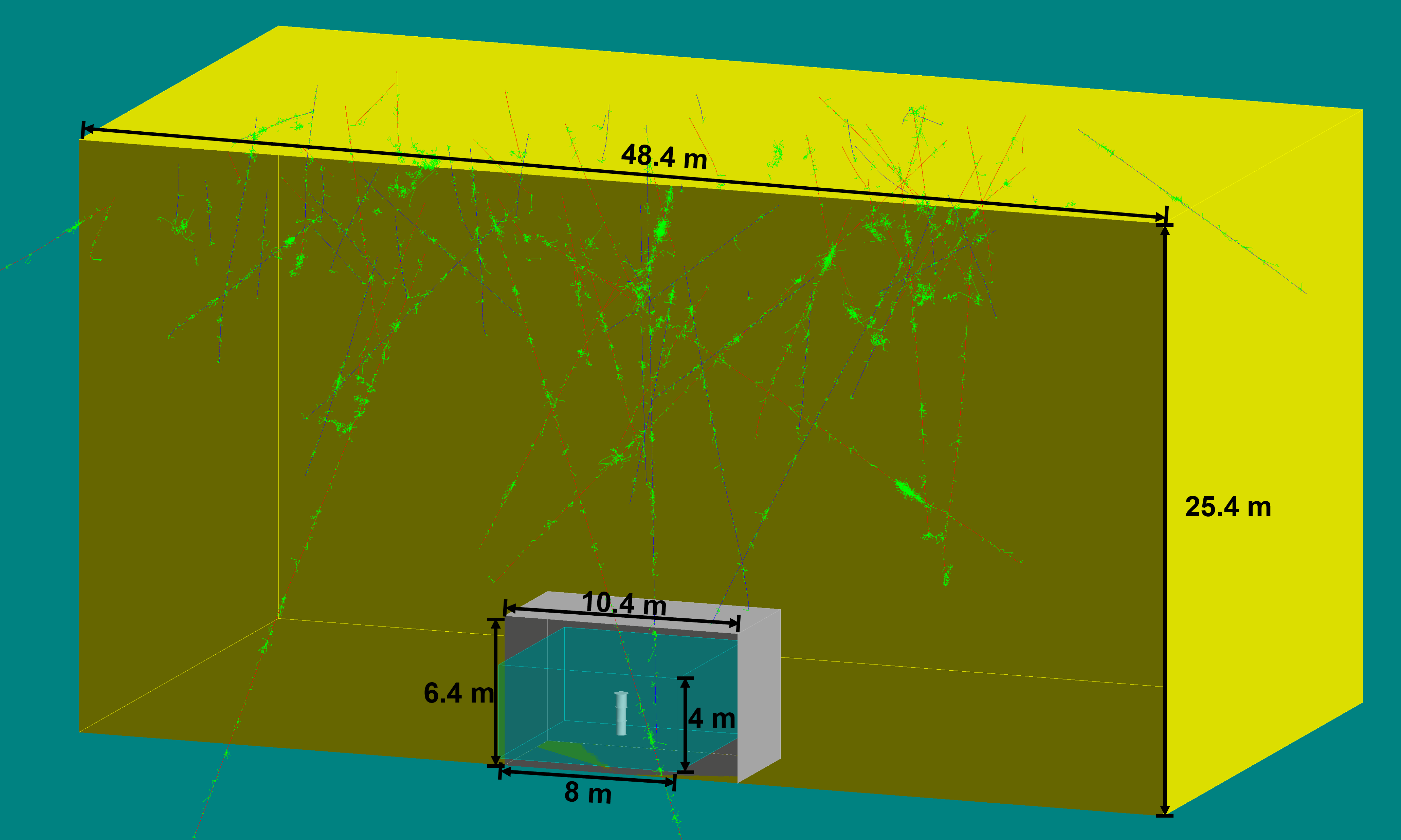}
    \caption{A \geant{} model cross-section rendering of a dilution fridge (cylinders) within a concrete cavity (grey) beneath a 19~m overburden (yellow) representing the PNNL Shallow Underground Laboratory (SUL). Some simulated cosmic-ray muon tracks are rendered in red and blue with tertiary ionized electron tracks shown in bright green.}
    \label{fig:underground}
    
\end{figure}

For the underground estimate, the simulated laboratory space is surrounded by 1.2~m thick concrete walls and a 19~m overburden (above and extending on all sides) composed of calcium carbonate (limestone) with a density of 2.8~g/cm$^3$. The simulation geometry is shown in Figure~\ref{fig:underground}. This simplified model does not include any vertical access shafts, near which there is significantly higher muon and neutron flux, nor does it account for the ``heaped'' profile of the SUL overburden, which affects the muon flux particularly at high zenith angles. We compensate for this lack of fidelity in the simulation model by normalizing to \textit{in situ} measurement. Only secondary muons are generated from a \SI{38}{m} square plane just at the top of the overburden (see Fig.~\ref{fig:underground}); all other cosmic-ray secondary particles contribute negligibly. To speed up the simulation, we employ a biasing technique where we immediately discard before propagation any initial muon whose momentum direction points more than \SI{3}{m} from the NaI detector. This biasing technique was compared to a full simulation, and the results agreed to within statistical uncertainty (as described in Section~\ref{sec:append-variance}).

\begin{figure}[ht!]
    \centering
    \includegraphics[width=0.85\textwidth]{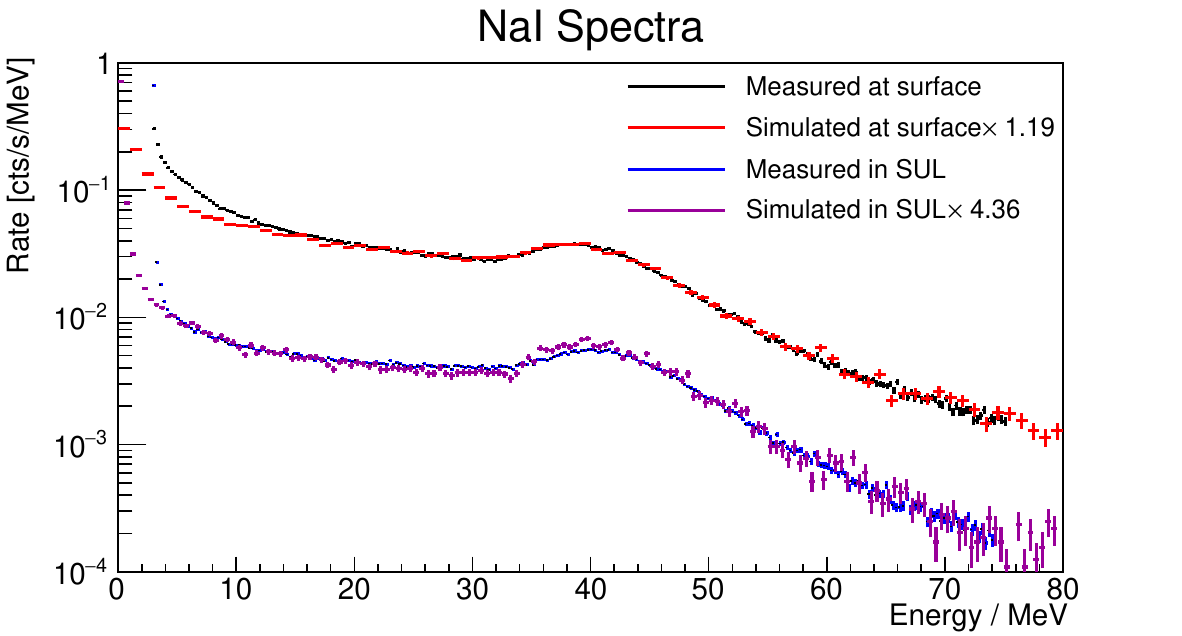}
    \caption{Comparison of cosmic-ray simulations to data collected from a \SI{3}{inch} NaI detector operated in the PNNL SUL with $\sim$19~m overburden (blue, bottom) and a surface laboratory (black, top).  The simulated spectra are linearly scaled horizontally and vertically to match the energy scale and amplitude, respectively. }
    \label{fig:muon_validation_SUL}
\end{figure}

Figure~\ref{fig:muon_validation_SUL} shows the measured data compared to the scaled simulation output. A simple two-parameter fit is performed to align the simulated data with measurement: a linear scale factor to convert from the arbitrary detector energy scale to energy deposited and a linear amplitude scale. The data measured at surface agree with the CRY-simulated rate within 20\%, which we consider reasonable agreement for comparison to an altitude, geomagnetic-latitude, and solar-cycle dependent estimate of the absolute rate of cosmic-ray secondaries at any specific location on the surface of the Earth. However, the rate measured underground is more than 4 times the simulated rate, most likely due to the simplified model with flat overburden and the proximity of the measurement detector to a vertical access shaft. 
The measured rate underground in the 30-50 MeV peak region is reduced by a factor of 6 relative to surface, in good agreement with prior measurements~\cite{aalseth2012}. The excess event rate in the measured spectra below 3~MeV is due to environmental gammas which are not included in the cosmic-ray simulations. The additional, smaller excess in the surface measurements up to $\sim$\SI{12}{MeV} are most likely due to details of the actual cosmic-ray secondary interactions with the surface laboratory building and detector effects such as quenching of neutron-induced scintillation, neither of which are modeled. 

To estimate the interaction rate and dose from cosmic-ray secondaries in superconducting qubit chips, we perform essentially the same procedure as for simulating the NaI detector. Particles are generated by \CRY{}, propagated by \geant{} through our model of a laboratory containing a dilution refrigerator and array of silicon chips (including walls and overburden for the SUL), and then normalized by the equivalent live time as reported by \CRY{} and scaled by the correction factors obtained from the NaI measurements. Respectively, the measurement determined normalization factors are 1.19 for the surface laboratory and 4.36 for the SUL. 

\subsection{Laboratory gamma flux}
\label{sec:envgammasource}
Estimating the radiation dose absorbed in superconducting devices from the ambient flux of gammas in the laboratory follows a similar procedure as for the cosmic-ray secondaries. We measure the spectrum of energy deposited in a radiation detector, simulate gammas from the laboratory walls interacting with the detector, and compare measurement to simulation to determine the absolute flux of gammas and to normalize subsequent simulations of superconducting devices. 

The measured gamma-ray spectrum was obtained with an unshielded high purity germanium (HPGe) detector. The Mirion GC14022 HPGe detector used for these measurements was a \textit{p}-type coaxial design with a vendor-specified 140\% relative efficiency at 1.33~MeV. The detector was placed near the center of one of the SUL labs, away from walls. Collected data were analyzed using PeakEasy v4.86 to identify key terrestrial background gamma-ray emitters. The most prominent isotopes (and their decay chain) identified were \iso{K}{40}~(K), \iso{Pb}{214}~(U), \iso{Bi}{214}~(U), \iso{Ac}{228}~(Th), \iso{Pb}{212}~(Th), \iso{Bi}{212}~(Th), and \iso{Tl}{208}~(Th), which accounted for 93.5\% of the total terrestrial gamma background spectrum.

For the simulation, the identified isotopes are distributing uniformly throughout a \SI{1.2}{m} concrete wall around the laboratory model and the characteristic gamma- and x-ray emissions are generated using \geant{}'s radioactive decay module~\cite{RDG6575161,RDG6575219}. The energy, position, and direction of each photon passing the surface of a 145\unit{cm} radius sphere (just large enough to contain the dilution refrigerator model) centered on the HPGe are recorded. An analysis, not presented here (see Ref.~\cite{pnnl-31996}), showed that the radiation emissions crossing the 145\unit{cm} radius sphere  have an angular distribution consistent with an isotropic flux. This permitted ``re-throwing'' (i.e., generation of new simulation primaries) of the gamma- and x-ray flux uniformly and isotropically from a smaller simulated sphere inside the room, substantially reducing the number of primaries required to parametrically explore specific shield design thicknesses and event rates at the location of the superconducting devices.

\begin{figure}
    \centering
    \includegraphics[width=\textwidth]{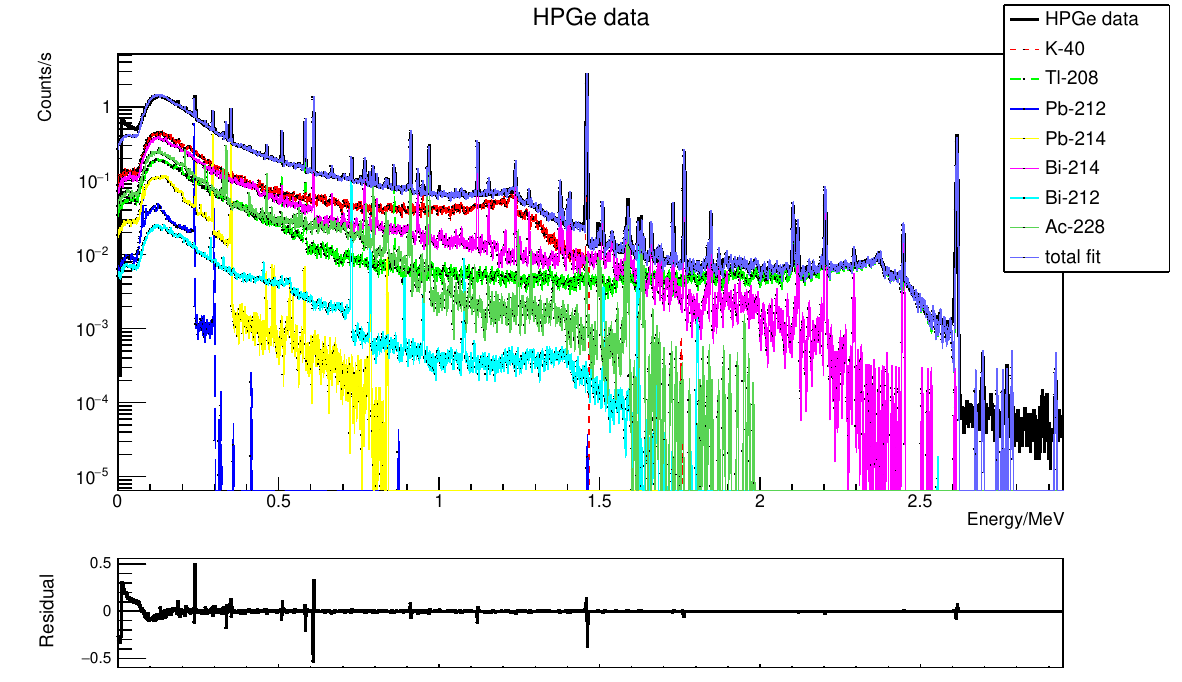}
    \includegraphics[width=\textwidth,height=0.6\textwidth]{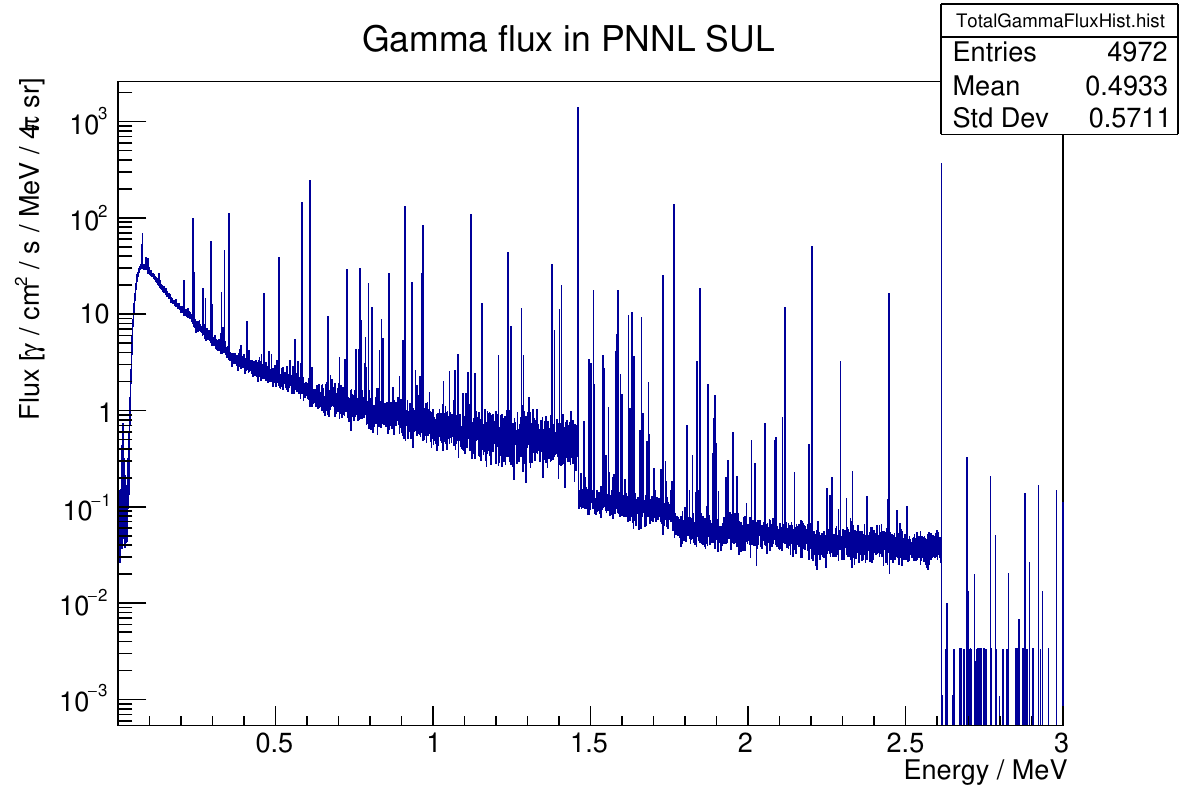}
    \caption{Top: spectrum recorded with an unshielded HPGe detector in the SUL and corresponding best fit produced from simulation data. Middle: Fit residual, normalized to the measured value at each energy. Bottom: The total flux of ambient gammas in the PNNL SUL estimated by the fit.}
    \label{fig:totalgammaflux}
\end{figure}

To determine the total gamma flux, separate simulations of the isotropic flux from each of the seven identified radionuclides are performed, and the energies deposited in a simulated HPGe detector are recorded.  The simulated data are then fit to the measured data to determine the relative amplitude of each component. Details of the fitting procedure are provided in Appendix~\ref{sec:hpgefits}. The fit to the measured data and the total gamma flux spectrum resulting from the fit are presented in Figure~\ref{fig:totalgammaflux}. The total integrated flux is approximately $\SI{7}{cm^{-2}s^{-1}(4\pi\ sr)^{-1}}$.  Although this flux spectrum is specific to the SUL, experience suggests that the ambient radiogenic gamma flux in most laboratories will be within a factor of a few of this result.  Measurements taken with a NaI detector in various labs at PNNL and MIT presented in Appendix~\ref{sec:compareNaI} support this assertion. 

\subsection{Internal sources of radiation}
To estimate the radiation dose from radioactive contaminants in materials inside the dilution refrigerator, we need two pieces of information for each component: (1) the total decay rate of radionuclides in that component and (2) the probability for a radioactive decay at the component's location to result in energy deposited into the superconducting quantum circuit device substrates, which we call the ``hit efficiency.'' To address the second requirement, we simulate radioactive decay emissions using our \geant{} simulation. The most common radioactive isotopes---namely \iso{U}{238}, \iso{Th}{232}, \iso{K}{40}, \iso{Co}{60}, \iso{Cs}{137} and  \iso{Pb}{210}---are distributed throughout each volume (excluding the silicon chip itself) in the simulation and allowed to decay via \geant{}'s Radioactive Decay Module. \iso{U}{238} and \iso{Th}{232} are allowed to decay through their entire chains (to stable \iso{Pb}{206} and \iso{Pb}{208} respectively), assuming secular equilibrium for simplicity. The resulting energy depositions in the silicon chips are recorded, and normalized to the silicon mass and total number of decays simulated. This creates the hit efficiency lookup table (Appendix~\ref{sec:append-hiteff}). The total interaction rate or dose for a given component is then equal to the component's mass times the differential contamination level (typically expressed in Bq/kg, i.e., decays per second per kilogram) times the entry in the corresponding table for the isotope of interest and the component's location. 

The level of radioactivity in different materials varies by several orders of magnitude, from 10's of Bq/kg in many common materials down to $\mu$Bq/kg in the purest materials such as silicon and OFHC copper. Different samples of the same material may have significant variation due to differences in manufacturing and handling, although we assume for simplicity that all instances of a given material have the same level of radioactivity. 
Thus, the radiation levels to which a device is exposed may vary substantially among laboratories and depend on the specific hardware and materials used in an experiment apparatus.
For this estimate. radioactivity measurements for most materials are selected from references in the literature that are representative of the most common fabrication processes. In the remainder of this section, we present radioactivity assay results for some key components and materials used in superconducting qubit experiments:  qubits, microwave printed circuit board (PCB) substrates, and cryogenic coaxial cables and connectors. The levels of radioactivity we have assumed for all materials considered are presented in Table~\ref{tab:radiopurity_lvls}. 

\subsubsection{Transmon qubit assay}


Superconducting qubits typically comprise crystalline dielectric substrates and superconducting thin films. For these devices, distributed circuit elements are constructed using standard nano- and micro-fabrication techniques. Josephson junctions (almost exclusively Al/AlOx/Al heterostructures) are formed using double-angle electron-beam shadow evaporation through a resist mask. In order to reduce energy loss at the microwave frequencies at which superconducting qubits (transmons in this case) operate, high-purity materials are consistently chosen and fabrication recipes are optimized to reduce process-induced contamination.
 
In this study, we screened transmon-based devices fabricated at MIT Lincoln Laboratory (MIT LL) for the presence of residual radioisotopes. Details of the fabrication process for these devices are presented in Appendix~\ref{sec:append-qubitfab}. 
Three substrates containing transmon qubit circuits 
were assayed for $^{232}$Th and $^{238}$U content at the PNNL’s Ultra-Low Background Detection facility, which specializes in highly sensitive radiopurity assays of materials~\cite{WOS:000592358200016,WOS:000504926900016,WOS:000348040900014,WOS:000596707400007,WOS:000395964400060,WOS:000518368800052}. Details of the measurement are reported in Appendix~\ref{sec:append-qubitassay}. 
\begin{table}
    \centering
    \caption{Assay results of the superconducting qubit chips and, for comparison, pure silicon and sapphire, which are commonly used for superconducting circuit substrates. The provided values for sapphire were measured on sapphire rods; we were unable to locate measurements for sapphire wafers.  The fabricated qubit chips have similar levels of radiopurity as pure substrates.}
    \label{tab:radiopurity_qubits}
    \vspace{0.5em}
    \begin{tabular}{l|cc|c}
    \toprule
    Sample & $^{232}$Th (mBq/kg) & $^{238}$U (mBq/kg) & Ref. \\
    \midrule
    Superconducting qubit chip  & 0.0065 $\pm$ 0.0012 & 0.014 $\pm$ 0.003 & This work \\
    Silicon & $<$0.0073 & $<$0.011 & \cite{Aguilar-Arevalo2021} \\
    Sapphire & 0.024$\pm$0.004 & $<$0.11 & \cite{nexocollaborationSensitivityDiscoveryPotential2018} \\
    \bottomrule
    \end{tabular}
\end{table}

Of the three transmon qubit chips assayed, only one replicate value was above the detection limit and is reported in Table~\ref{tab:radiopurity_qubits}. The error on the single measurement is the instrumental error. Detection limits were 0.003 and 0.009~mBq/kg for $^{232}$Th and $^{238}$U, respectively. For comparison, we also provide in Table~\ref{tab:radiopurity_qubits} assay results of pure silicon and sapphire found in literature.


The measured activity levels of the assayed devices are similar to those of pure silicon, which indicates that neither the transmon qubit fabrication process, nor the additional materials applied in the circuitry, elevate the trace levels of $^{232}$Th and $^{238}$U above the purity of the substrate. The qubit chip radiopurity is comparable to or better than OFHC copper, commonly used in device packages, which can be one of the purest materials available commercially, with uranium and thorium levels typically ranging from $\sim$0.1-50~$\mu$Bq/kg~\cite{abgrall2016, aprile2011a, aprile2017a}. 
In contrast, external materials such as the surrounding readout wiring can have as much as one to six orders of magnitude higher activity than the qubits themselves.

\subsubsection{Interposer laminate assay}
Frequently, qubits are packaged with an ``interposer:'' a printed circuit board (PCB) to which the qubit device is connected by wirebonds and which brings the RF and DC connections for the qubit to discrete connectors outside the package. The bulk material of the interposer is often a composite dielectric. The interposer may have metallization layers that are gold, copper, tin, or aluminum. Typical choices for the interposer bulk material include Rogers TMM10, Rogers ``RO~series'', and alumina, although more radiopure materials such as silicon and sapphire may also be used. As a general rule, ceramics often have above-average levels of radioactivity. This, coupled with their close proximity to the qubit, suggests that the interposer may be a significant source of radiation events in the chip. In Table~\ref{tab:hpge_pcb_results}, we report the radionuclide assay of samples of Rogers TMM10 and RO4350B PCB substrates, each counted for approximately 5 days in a high purity germanium (HPGe) system. 
As can be seen in Table~\ref{tab:radiopurity_lvls}, these ceramic substrates do indeed have substantially higher levels of radioactivity than other materials in the dilution refrigerator. FR4, another common PCB substrate, has similarly high levels of radioactivity~\cite{cardani2023, ARMENGAUD20131}.

\begin{table}
    \centering
    \caption{Radioactive isotopes measured in Bq/kg in Rogers ceramic PCBs by HPGe counting. Only isotopes identifed with $>$90\% confidence are reported. The uncertainty in the final digit is reported in parentheses.}
    \label{tab:hpge_pcb_results}
    \sisetup{table-format=2.1(1)}
    \begin{tabular}{lr|SSSSSSS}
        \toprule
        Sample & Mass & {\iso{K}{40}} & {\iso{Tl}{208}} & {\iso{Pb}{212}} & {\iso{Bi}{214}} & {\iso{Pb}{214}} & {\iso{Ra}{226}} & {\iso{Pb}{210}} \\ 
        \midrule
         TMM10 & \SI{200}{g} & 17.3 \pm 0.9 & 1.51 \pm 0.06 & 5.5 \pm 0.3 & 28.9 \pm 0.4 & 25.4 \pm 0.8 & 29 \pm 2 & {-} \\
         RO4350B & \SI{30}{g} & 9.1 \pm 0.8 & 4.9 \pm 0.2 & 15.1 \pm 0.9 & {-} & 11.2 \pm 0.4 & 8 \pm 4 & 11 \pm 2 \\
        \bottomrule
    \end{tabular}
\end{table}

\subsubsection{SMA connector and coaxial cable assay}
Because superconducting qubits are generally sensitive to magnetic fields, non-magnetic materials, such as BeCu, are often selected for the package and nearby interconnects. Device packages often have non-magnetic (BeCu) coaxial connectors attached to the package perimeter. The coaxial cables that mate to these package connectors are also selected as non-magnetic. The proximity and relatively high activity of BeCu may produce a significant interaction rate in the nearby superconducting devices. We assayed two SMA connectors, one from each of two cables supplied by Bluefors and from Crystek, both \SI{50}{Ohm}, \SI{0.086}{inch} (\SI{2.2}{mm}) OD hand-formable coaxial cable. We also assayed three sections of the Crystek cable with braided tinned copper outer conductor. Measurements followed the dissolution and ICP-MS technique described in Appendix~\ref{sec:append-qubitassay}. Before dissolution, the cables and connectors were cleaned by sonication at room temperature, first with a 2\% Micro90 solution, then MilliQ water, for 15 minutes each with a triple rinse in MilliQ water after each sonication step. Only the metals in each part were dissolved in a mixture of \ce{HNO_3} and \ce{HCl} and sampled. The results reported in Table~\ref{tab:assay_sma} are normalized only to the sampled mass fraction. The remaining materials were primarily the PTFE dielectric and silicone o-ring in the connector, both of which materials are typically low in radioactivity. As Table~\ref{tab:assay_sma} shows, the body of the SMA connector has high levels of \iso{U}{238} and \iso{Th}{232}, most likely due to BeCu, while the cable body is much more radiopure.

\begin{table}
    \centering
    \caption{ICP-MS assay results for SMA connectors and hand-flex 086 coaxial cable. Reported uncertainties are given by the instrumental precision. Detection limits are calulated as three times the standard deviation of process blanks. }
    \label{tab:assay_sma}
    \begin{tabular}{lc|ccc|cc}
    \toprule
    Sample &  & Total Sample & Assayed & Mass Fraction & \iso{Th}{232} & \iso{U}{238} \\
           &  & Mass (g)     & Mass (g) & Assayed      & (mBq/kg)         & (mBq/kg) \\
    \midrule
        connector & 1 & 2.9040 & 2.6336 & 0.907 & $1430 \pm \phantom{0}20$ & $21000 \pm 2000$ \\
                  & 2 & 2.8953 & 2.6432 & 0.913 & $2240 \pm 140$ & $25000 \pm 2000$ \\
    \midrule
         cable & 1 & 0.1429 & 0.1056 & 0.739 & $<0.130$ & $<0.39$ \\
               & 2 & 0.1872 & 0.1334 & 0.713 & $<0.152$ & $<0.42$ \\
               & 3 & 0.1552 & 0.1111 & 0.716 & $<0.16\phantom{0}$ & $<0.49$ \\
    \bottomrule
    \end{tabular}
\end{table}

\ctable[caption={Radio-contaminant levels of materials and small parts used to estimate the ionizing radiation interaction rate and dose generated by materials inside the dilution refrigerator. Where a range of reported values exists, we choose values roughly consistent with the distribution median. },
        label={tab:radiopurity_lvls},
        mincapwidth=\textwidth,
        super, 
        ]
    {l|ccccccc|c}
    {
    \\
        \tnote[a]{\iso{Pb}{210} is not typically measured, however there is evidence that commercial OFHC copper contains bulk \iso{Pb}{210} contamination at the $\sim$10's of \unit{mBq/kg} level, several orders of magnitude out of equilibrium with ancestor \iso{U}{238}~\cite{abe2018, balogh2021}.}
        \tnote[b]{Copper and its alloys are assumed to have the following radioactive isotopes from cosmogenic activation in equilibrium (rates in \unit{mBq/kg}): \iso{Co}{60} (2.1), \iso{Fe}{59} (0.5), \iso{Co}{58} (1.7), \iso{Co}{57} (1.8), \iso{Co}{56} (0.2), \iso{Mn}{54} (0.2), \iso{V}{48} (0.1), \iso{Sc}{46} (0.05) \protect{\cite{laubenstein2009}. These rates are appropriate for sea level altitude and will increase at higher elevation. Other materials, such as mumetal which is primarily nickel, may also have significant rates of activation-maintained radioactivity but are not modeled here.}}
    }
    {
    \toprule
    & \multicolumn{7}{c|}{Isotope concentrations (mBq/kg)} & \\
    Material  & \iso{U}{238} & \iso{Th}{232} & \iso{K}{40} & \iso{Co}{60} & \iso{Cs}{137} & \iso{Pb}{210}\tmark[a] & Act.\tmark[b] & Ref. \\ \hline
    copper & 0.070 & 0.021 & 0.023 & 0.002 & - & 40 & 6.6 & \cite{aprile2011a, abe2018, laubenstein2009} \\
    lead & 0.04 & 0.005 & 0.1 & - & - & 200000 & - & \cite{abgrall2016, Orrell20161271, Keillor2017185} \\
    steel & 130 & 2.4 & 10 & 8.5 & 0.9 & - & - & \cite{aprile2011a} \\ 
    aluminum & 66 & 200 & 2100 & - & - & - & - & \cite{aprile2011a} \\
    gold & 74 & 19 & 150 & - & - & - & - &  \cite{leonard2008a, abgrall2016} \\
    brass & 4.9 & 3.5 & 40 & - & 2.6 & 40 & 6.6 & \cite{ARMENGAUD20131, armengaud2017} \\ 
    Kapton & 10 & 20 & 60 & 3 & - & - & - & \cite{aprile2017a, armengaud2017} \\
    Al bonding wire & 110 & 370 & 100 & - & - & - & - & \cite{abgrall2016} \\
    mumetal & 20 & 7 & 15 & - & - & - & - & \cite{ILIAS} \\
    isolator & 240 & 190 & 2000 & - & 50 & - & - & \cite{cardani2023} \\
    HEMT & 1000 & 890 & 10000 & - & 210  & - & - & \cite{cardani2023} \\
    K\&L filter & 9 & 23 & 100 & 5 & 1.9  & - & - & \cite{cardani2023} \\
    attenuator & 200 & 52 & 140 & - & 13   & - & - & \cite{cardani2023} \\
    alumina & 5000 & 66 & 600 & - & - & - & - & \cite{ILIAS} \\
    Rogers TMM10 & 29000 & 5500 & 17000 & - & - & - & - & this work \\
    Rogers RO4350B & 11000 & 15000 & 9000 & - & - & - & - & this work \\
    SMA connector & 23000 & 1800 & - & - & - & - & - & this work \\
    coaxial cable & 0.4 & 0.15 & - & - & - & - & - & this work \\
    qubit chip & 0.014 & 0.0065 & - & - & - & - & - & this work \\
    \midrule
    Indium & \multicolumn{2}{c}{\iso{In}{115}: 250000} &  &  &  &  &  &  \\

    \bottomrule
    
    }
\clearpage
\subsection{Total radiation budget}
\begin{figure}
    \centering
    \includegraphics[width=0.98\textwidth]{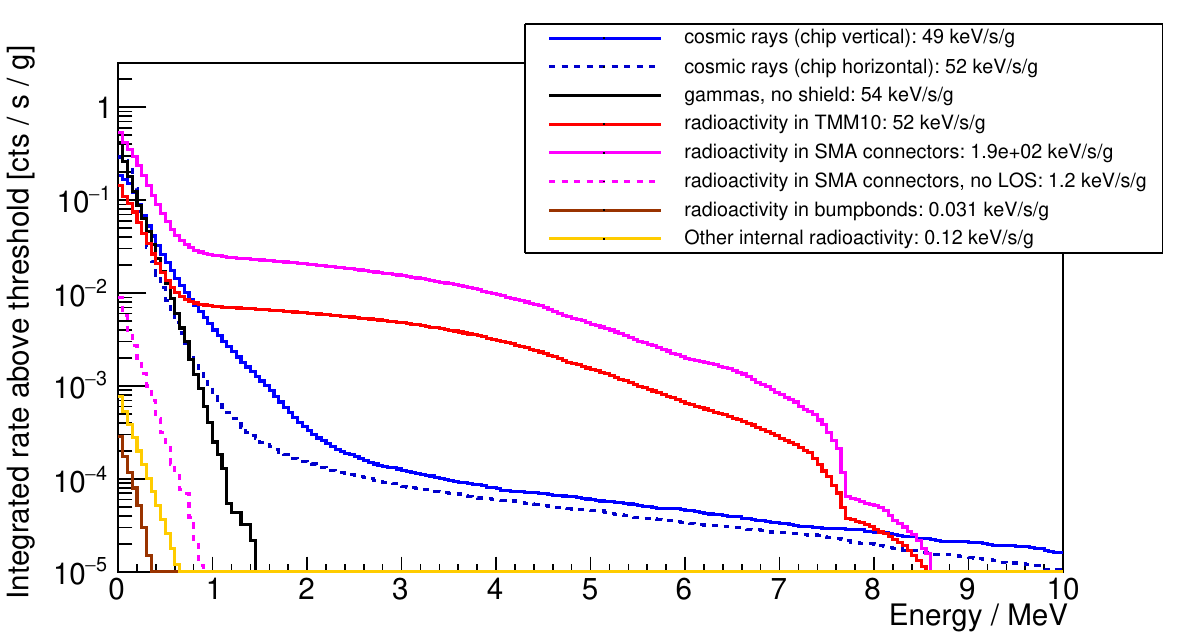}
    \includegraphics[height=0.8\textwidth]{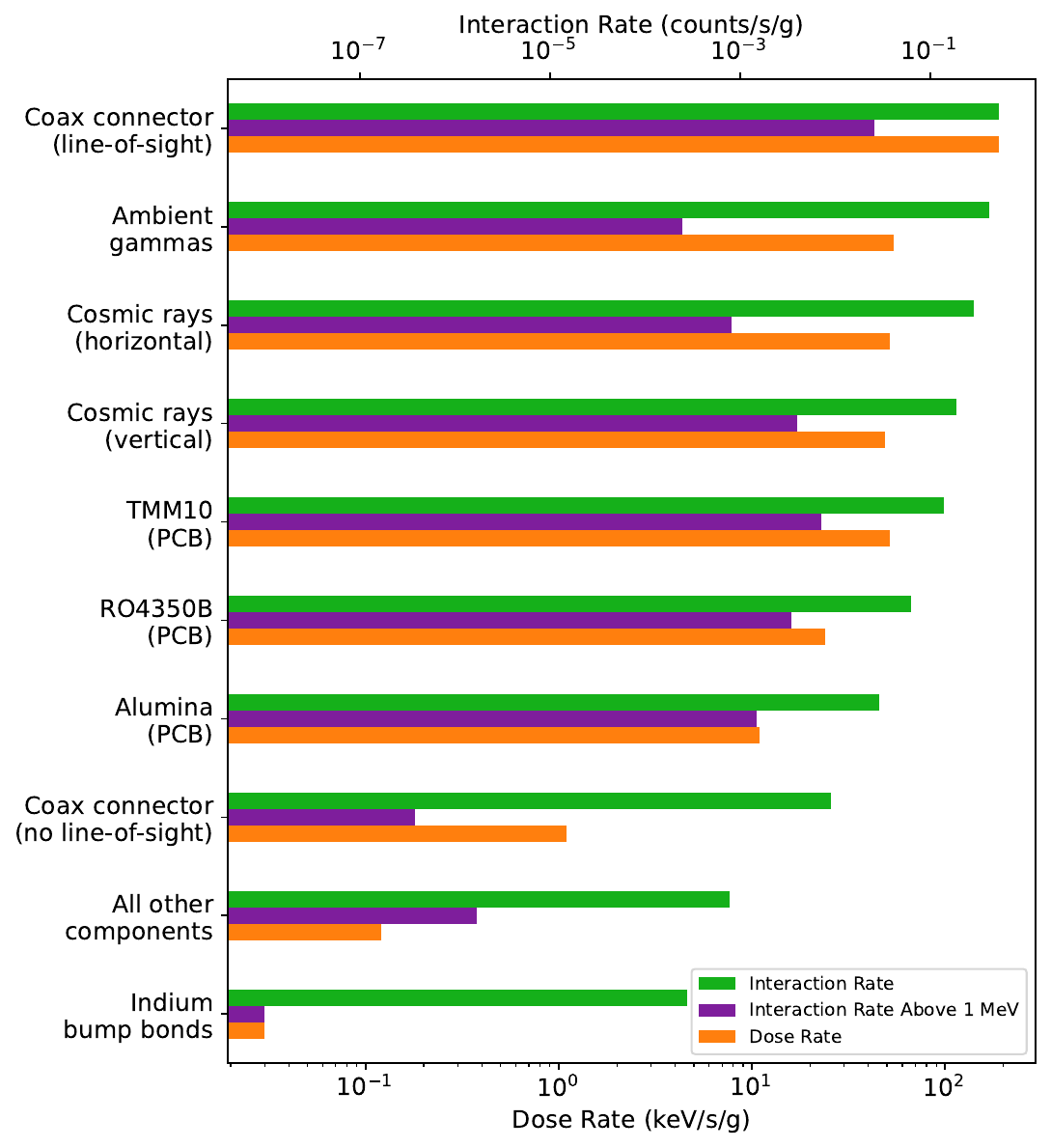}
    \caption{Top: Total rate of ionizing radiation events depositing energy greater than a threshold (horizontal axis) in a silicon substrate inside a dilution refrigerator operating at sea level. Bottom: Bar chart summary of spectra comparing total interaction rate, interaction rate of events depositing $>$\SI{1}{MeV}, and average dose rate (total interaction rate times the average energy deposited per interaction). }
    \label{fig:radiation_spectrum}
\end{figure}

Table~\ref{tab:internalrate} lists the average interaction rates and doses from different sources of ionizing radiation for a typical superconducting device in a laboratory near sea level. Figure~\ref{fig:radiation_spectrum} (top) shows the integrated interaction rate above a given energy threshold. Figure~\ref{fig:radiation_spectrum} (bottom) summarizes the contributions from each major source in a bar chart. The rate of interactions depositing greater than \SI{1}{MeV} of energy into the chip substrate is much lower than the total rate for sources without line-of-sight to the device. Excluding line-of-sight sources, the total interaction rate is driven roughly equally by the ambient gamma flux and by cosmic rays. In most cases, the ratio between the dose rate and the interaction rate is roughly constant across sources, with the principle exceptions of cosmic rays and indium bump bonds. Indium emits only relatively low energy betas, unlike other radio-contaminants that emit gammas. The dose rate for cosmic rays is largely independent of orientation, whereas the interaction rates, both total and greater than \SI{1}{MeV}, vary with orientation. A simple test comparing the induced error rates of a superconducting chip in horizontal versus vertical orientation may shed light on whether errors scale with interaction rate or with dose, which would provide insight on the possible underlying mechanisms. 

\begin{table}
    \centering
    \caption{Simulated interaction rate for all events and for events that deposit greater than \SI{1}{MeV} in a silicon chip from environmental gammas, cosmic rays, and trace radioactivity inside an unshielded dilution refrigerator at sea level. Upper limits due to finite simulation statistics are reported at 90\% confidence level. 
    }
    \label{tab:internalrate}
    \sisetup{separate-uncertainty, table-align-uncertainty=false, table-align-exponent=true}
    \begin{tabular}{|llc|S[table-format=3.1e-1]|S[table-format=3.1e-1]|}
        \multicolumn{5}{l}{\vspace{-0.75em}} \\
        \hline
        Component & Material & {Mass} & {Interaction rate} & {Rate $>$1 MeV} \\
        & & {(kg)} & {($10^{-3}$~counts/s/g)} & {($10^{-3}$~counts/s/g)} \\
        \hline
        \multicolumn{3}{|l|}{Cosmic rays (chip horizontal)} & 290 & 0.81 \\
        \multicolumn{3}{|l|}{Cosmic rays (chip vertical)} & 190 & 4.0 \\
        \multicolumn{3}{|l|}{Ambient Gammas} & 420 & {$<$}0.025 \\
        \multicolumn{3}{|l|}{Ceramic PCB interposers} & & \\
        \quad  & alumina & \SI{780}{mg} & 29 & 1.5 \\
        \quad  & RO4350B & \SI{370}{mg} & 63 & 3.5 \\
        \quad  & TMM10   & \SI{550}{mg} & 140 & 7.2 \\
        \multicolumn{3}{|l|}{Coax connectors on package} &  &  \\
        \quad inside (line-of-sight) & SMA & {$10\times\SI{2.3}{g}$} & 530 & 26\\
        \quad outside (no line-of-sight) & SMA & {$10\times\SI{2.3}{g}$} & 8.9 & 3.8e-4 \\
        Bump bonds & indium &{\SI{20}{\mu g}} & 0.28 & {$<$}1e-5 \\
        \multicolumn{3}{|l|}{All other components (itemized below)} & 0.72 & 0.0017 \\
        \hline
        \multicolumn{3}{|l|}{\quad Fridge stages and shields} & 0.23 & 4.4e-5 \\
        \quad \quad MXC stage & Cu & 4.6 & 0.0027  & 3.9e-7 \\
        \quad \quad CP stage & Cu & 3.3 & 2.9e-4 & 1.5e-7 \\
        \quad \quad Still stage & Cu & 5.9 & 2.1e-4 & 2.1e-8 \\
        \quad \quad 4K stage & Cu & 8.7 & 8.6e-5 & 1.5e-8\\
        \quad \quad 50K stage & Cu & 5.1 & 1.6e-5 & {$<$}2e-6\\
        \quad \quad Vacuum flange & steel & 21 & 7.9e-4 & {$<$}4e-5\\
        \quad \quad Still can & Cu & 6.3 & 0.0019 & 2.9e-7\\
        \quad \quad 4K can & Al & 4.1 & 0.058 & 4.2e-6\\
        \quad \quad 50K can & Al & 5.7 & 0.047 & 2.5e-5\\
        \quad \quad Vacuum can & Al & 21 & 0.11 & 1.2e-5 \\
        \quad \quad Gold plating & gold & 0.5 & 0.010 & 1.7e-6\\
        \hline
        \multicolumn{3}{|l|}{\quad Experiment readout} & 0.49 & 0.0017 \\
        \quad \quad Wirebonds & Al/Si & {$10\times\SI{0.1}{mg}$} & 0.0029 & 1.8e-4 \\
        \quad \quad Package & Cu & 0.1 & 0.042 & 0.0013 \\
        \quad \quad Package Fasteners & brass & {$10\times\SI{0.3}{g}$} & 0.0045 & 8.9e-5 \\
        \quad \quad Cryo filters & K\&L & {$10\times\SI{15}{g}$} & 0.10 & 2.8e-5\\
        \quad \quad Closest coax cable & semirigid & {$10\times\SI{10}{cm}$} & 4.5e-6 & {$<$}9e-9 \\
        \quad \quad Coldfinger & Cu & 1.8 & 0.0065 & 1.1e-6 \\
        \quad \quad Inner shield &  &  & 0.11 & 1.4e-7 \\
        & Cu & 1 & 9.9e-4 & 1.4e-7\\
        & Al & 1 & 0.098 & {$<$}3e-3 \\
        & mumetal & 1 & 0.0065 & {$<$}4e-5 \\
        \quad \quad MXC DC feedthroughs & BeCu & {100 pins} & 4.0e-5 & 9.2e-9 \\
        \quad \quad MXC RF feedthroughs & SMA & {$10\times\SI{2.3}{g}$} & 0.082 & 4.0e-5 \\
        \quad \quad MXC RF attenuators & & {$10\times\SI{5}{g}$} & 0.0018 & 7.9e-7\\
        \quad \quad MXC isolators & & {$10\times\SI{145}{g}$} & 0.14 & 3.9e-5 \\
        \quad \quad 4K HEMT amplifiers & & {$10\times\SI{17}{g}$} & 8.6e-4 & {$<$}2e-5\\
        
        \hline
    \end{tabular}
\end{table}

Typical variation across different locations will be within a factor of $\sim$3 for both the ambient gamma flux (see Appendix~\ref{sec:compareNaI}) and for cosmic-ray muons~\cite{grieder2001}.  The level of radioactive contamination in materials inside the dilution refrigerator may easily vary by one to two orders of magnitude compared to our estimates. 
However, to be significant compared to rates from the ambient gamma and muon fluxes, internal sources would need to be either massive (kg) or very close to the device (cm).  Several interesting conclusions can be drawn, as described in the next three paragraphs.

\paragraph{Ambient gammas and cosmic-ray muons contribute roughly equally.} This statement holds if the effective threshold energy is below $\sim$\SI{0.5}{MeV}. Each of these sources produces an event rate on the order of 0.3~counts/s/g, or around 1 count/minute for a cm$^2$ chip. If the rate of ionizing radiation events must be reduced for a device, both shielding (for ambient gammas) and operating in an underground facility (for cosmogenic muons) are required. In an above-ground laboratory, shielding alone would reduce the overall rate by no more than about 70\%. 

\paragraph{Nearby ceramics, BeCu, and indium dominate internal sources.} Although only roughly a gram in mass, ceramic PCB interposers and BeCu SMA connectors contribute over two orders of magnitude more than all other internal components combined. This is due to their relatively high radioactivity content and close proximity to devices. This situation is exemplified when we consider two different locations for SMA connectors, either directly outside the copper package or inside with direct line-of-sight to the device of interest. In the latter case, the contributed event rate (assuming near-zero threshold) is over 20 times higher. SMA connectors located at the mixing chamber (MXC) plate contribute 100 times less. We note that the pieces assayed were male SMA connectors on cables, whereas a typical part with line-of-sight to the device would be the inner pin of a bulkhead solder receptacle, which may have significantly lower radioactivity. On the other hand, the inner pin is usually BeCu, which is the most likely source of contamination in the male SMA connector. These two sources also contribute at similar levels to the environmental sources, which suggests that any shielded and underground operation will benefit only marginally unless these materials are also addressed. Indium has by far the highest specific activity of all materials considered in this study and the closest proximity to the chips. Only the extremely small mass (10's of $\mu$g) prevent it from being a more significant contributor. We estimate that indium bump bonds contribute 3$\times10^{-4}$ counts/s/g ($\sim$1 count/10~hours for a cm$^2$ chip) to the total interaction rate, which is comparable to the sum of all other internal components considered excluding ceramic and BeCu parts (7$\times10^{-4}$ counts/s/g).

\paragraph{Alphas and cosmic-ray secondaries can produce high energy events, unlike gammas.}  In Table~\ref{tab:internalrate}, when enumerating the ``total interaction rate,'' we consider all interactions that inject energy above the silicon bandgap energy. Data from microcalorimeter detectors instrumented with multiple sensors~\cite{agnese2018} and from energy transport simulations~\cite{kelsey2023} suggest that this could be enough energy to produce a phonon cloud subsequently filling the entire substrate and therefore have high probability to be detected (or generate an error in a superconducting qubit). However, if the phonons are efficiently absorbed or converted to low energy, (\textit{e.g.}, by superconducting or normal-metal ground planes~\cite{riwar2016,iaia2022}), then the size of the phonon cloud and the probability to reach the active device elements may exhibit some energy dependence. A 5 MeV energy deposit will raise the temperature of a \SI{10}{mm^3} silicon chip to $\sim\SI{150}{mK}$, which may affect devices regardless of how effectively the initial athermal phonon population is downshifted. If there is a significant threshold effect, gammas become much less of a concern, and the error-inducing event rate will be dominated almost entirely by line-of-sight alpha emission. High energy alpha, proton, and neutron interactions can also produce dislocations in the crystal, which may, for example, affect local two-level systems over long timescales~\cite{thorbeck2023}.

\section{Abatement of ionizing radiation}
In this section, we consider steps to reduce the rate of ionizing radiation interactions within superconducting devices. The three major background contributors (cosmic-ray muons, external gammas, and internal contamination) must be reduced with different methods: shielding cosmic-ray particles with overburden (going underground), gamma shielding (usually lead or tungsten), and replacement of relevant materials with lower-radioactivity alternatives. We begin by describing the design and predicted ionizing radiation rates in the Low Background Cryogenic Facility (LBCF) at PNNL: a dilution refrigerator operating in PNNL's Shallow Underground Laboratory (SUL) outfitted with a lead gamma shield. In Section~\ref{sec:devices_in_lbcf} we predict how superconducting qubit devices might perform in the reduced radiation environment of the LBCF. In Section~\ref{sec:further_reduction}, conclude with a discussion of how one would further reduce the rate of ionizing radiation-induced interactions in superconducting devices, eventually adopting techniques used in ultra low background experiments such as those searching for dark matter, which target ionizing interaction rates on the order of 1 event per gram per month~\cite{agnese2017}. 

\subsection{The Low Background Cryogenic Facility (LBCF)}
The LBCF is designed to enable the study of superconducting device performance in a low ionizing radiation environment, limited by the residual cosmic-ray muon flux in the PNNL SUL. A Bluefors LD-400 dilution refrigerator has been operating in the SUL space since 2023. 
The SUL is described in Ref.~\cite{aalseth2012}. The entire SUL including the LBCF laboratory is operated as a class 10,000 or better cleanroom with focus on controlling radioactivity-bearing particulates. The 30~m.w.e overburden reduces the cosmic-ray muon flux by a factor of $\sim$6 and the cosmic-ray neutron and proton fluxes by $>$100~\cite{aalseth2012}. Our simple simulation model is described in Section~\ref{sec:cosmicraysource} and depicted in Figure~\ref{fig:underground}. In addition to the overall reduction in total muon flux, the muon angular distribution is slightly more downward-going than at surface, which has a small effect on the relative interaction rates for horizontally- versus vertically-oriented chips. The interaction rate for a \SI{12.5}{mm^2} chip in the SUL oriented horizontally is roughly twice that for a vertically oriented chip, compared to a ratio of $\sim$1.5 at surface. The muon spectrum underground also has higher average energy because lower-energy muons are attenuated more efficiently, but this does not significantly affect the results. 

The residual muon flux determines the required efficiency of the gamma shield: reducing the gamma-induced rate below a few percent of the muon-induced rate is unproductive. From Table~\ref{tab:internalrate}, the optimal residual gamma rate, accounting for the factor 6 reduction in muon flux, is $\sim$0.001~to~0.01~counts/s/g, or a reduction factor of $\sim$100 to 1000. 

Designing the lead gamma shield proceeds in two stages. First, we simulate a 4$\pi$ fully-enclosing shield of varying thickness around the dilution refrigerator and record the residual rate from environmental gammas to determine the required thickness. Then we introduce gaps into the the shield model to account for practical considerations such as mechanical supports and interfaces, penetrations for cooling and signal lines, and gaps from finite mechanical tolerances, and evaluate how these modifications reduce the effective shielding rate. 

To simplify the large possible parameter space for the specific shield design, we assume the shield is built primarily from ``standard'' \dims{2}{4}{8}\unit{in^3} lead bricks. This sets a natural step size (\SI{2}{in}) for considering variations in the shield thickness. We also use this assumption to simplify modeling gaps in the shield to set extremely conservative tolerances (i.e., the possibility for gaps between lead shield bricks). In the simulation, bricks are arranged with an edge parallel to the shield's thickness, such that any gaps between bricks become direct holes in the shield. We then add gaps of the specified tolerance around each individual brick and evaluate the effect. Our simulations suggest that gaps up to 1/8\unit{in} ($\sim$\SI{3}{mm}) thickness have negligible impact on shielding performance. In the actual design, the bricks are arranged with overlapping gaps to prevent such direct lines of sight, and the only full gaps would be at the seams where movable sections meet. These full gaps are mitigated in the design with stepped faces at the seams. 

\begin{figure}
    \centering
    \includegraphics[width=0.9\textwidth]{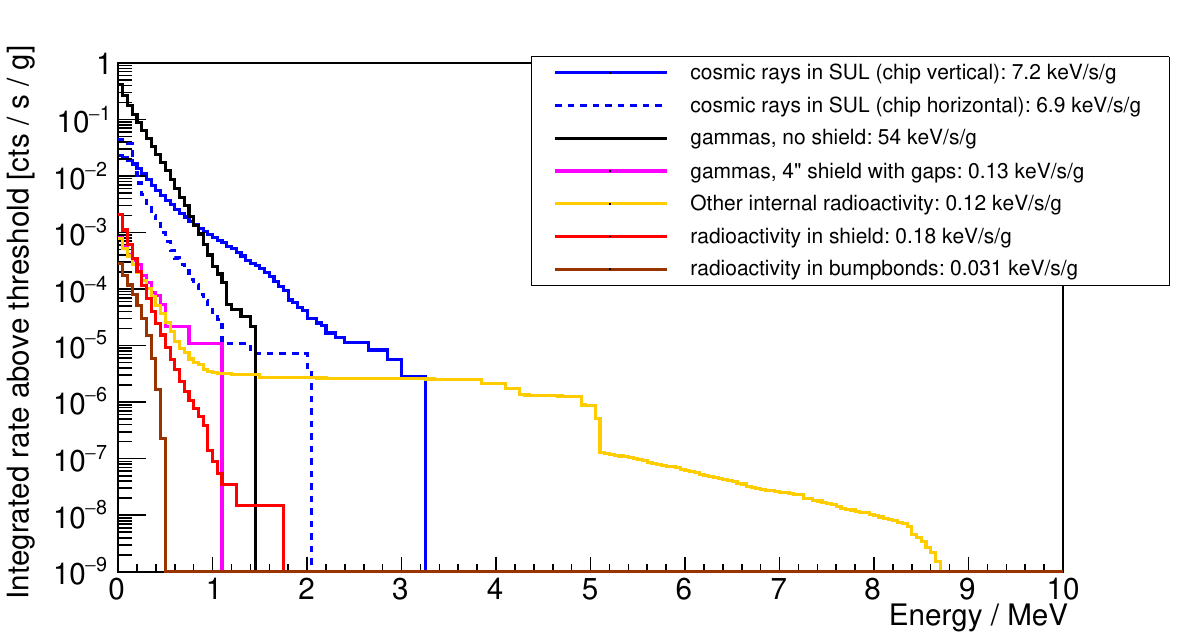}
    \caption{Simulated rate of ionization interactions depositing energy greater than a threshold (horizontal axis) in silicon substrates located in the PNNL SUL, for cosmic rays, ambient gammas, residual gammas for a 4" thick lead shield with penetrations (for mechanical supports, vacuum and helium connections, and instrumentation), radioactivity inside the dilution refrigerator, and radioactivity from the shield itself. 
    }
    \label{fig:specvsshieldthick}
\end{figure}

\begin{table}
    \centering
    \caption{Simulated interaction rate and dose in silicon chips from cosmic-ray and ambient gamma-ray sources vs. various design configurations. The final detailed shield includes holes in the shield for cryogenic and power service (see text body). 
    }
    \sisetup{table-format=3.2e-1}
    \begin{tabular}{l|S[table-format=2.3]SS}
        \toprule
        Source, Configuration & {Interaction rate} & {Rate $>$1 MeV} & {Dose rate} \\
        & {(10$^{-3}$ cts/s/g)} & {(10$^{-3}$ cts/s/g)} & {(keV/s/g)} \\
        \midrule
        cosmic-ray muons in SUL \\
        \quad Chip vertical & 22 & 1.0 & 7.4 \\
        \quad Chip horizontal & 44 & 0.033 & 6.9 \\
        Residual environmental gammas \\
        \quad 2'' (5.08~cm) enclosed lead shield & 11 & 0.033 & 2 \\
        \quad 4'' (10.16~cm) enclosed lead shield & 0.54 & {$<$}0.03 & 0.10 \\
        \quad 6'' (15.24~cm) enclosed lead shield & 0.043 & {$<$}0.03 & 0.010 \\
        \quad 4'' (10.16~cm) lead shield with gaps & 0.80 & 0.011 & 0.13 \\
        Internal radioactivity & 0.72 & 0.0017 & 0.11 \\
        Gammas from shield & 2.1 & 9.0e-5 & 0.18 \\
        \quad \iso{Pb}{210} in lead & 1.7 & 1.5e-7 & 0.12 \\
        \quad Aluminum support & 0.41 & 9.0e-5 & 0.05 \\
        \bottomrule
    \end{tabular}
    
    \label{tab:ratevsshieldthick}
\end{table}

Figure~\ref{fig:specvsshieldthick} shows the simulated rate of ionizing interactions above a set threshold in the silicon substrates for cosmic rays and for the unshielded and residual ambient gamma flux. Table~\ref{tab:ratevsshieldthick} summarizes the results and additionally presents residual gamma-induced rates for fully-enclosed ideal shields of varying thickness. As discussed previously, the details of the overburden profile affect the cosmic-ray angular distribution and therefore the spectrum in the silicon. In particular, more flux at higher zenith angle (closer to the horizon) would lead to higher average energy deposition for horizontally-oriented chips and conversely lower average energy deposited for vertically-oriented chips. This is likely the case in the PNNL SUL because we used a flat overburden in our simulation unlike the true heaped profile, The total ionizing radiation interaction rate decreases with additional shielding, but with significantly diminishing returns as the residual gamma rate becomes much less than the cosmic-ray rate. With 4\unit{inches} of lead shielding, the simulated gamma rate reaches about 2\% of the cosmic-ray rate for a vertically-oriented chip.

\begin{figure}
    \centering
    \includegraphics[width=0.48\textwidth, trim=300 20 300 20, clip]{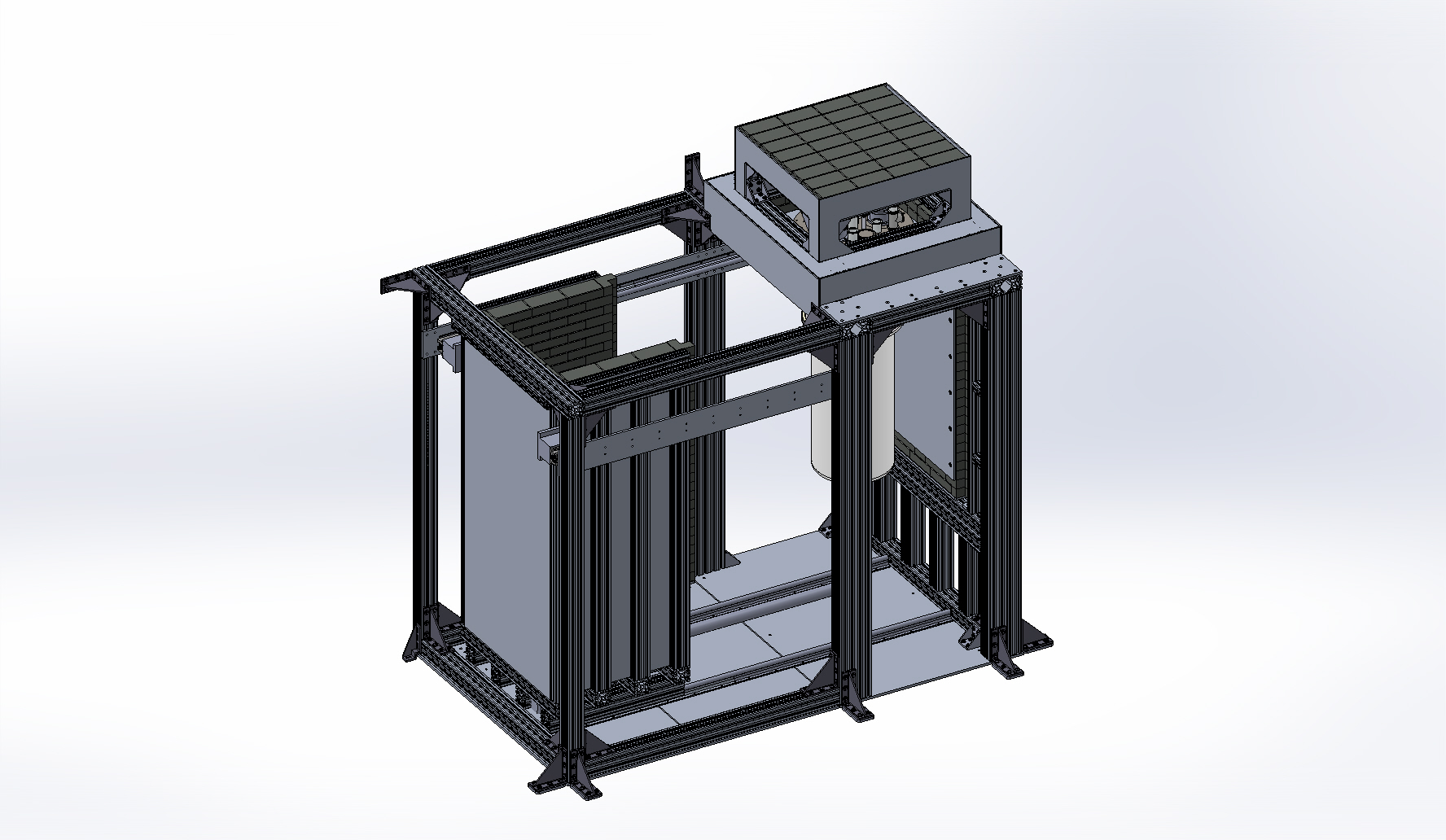}
    \includegraphics[width=0.48\textwidth, trim=300 20 300 20, clip]{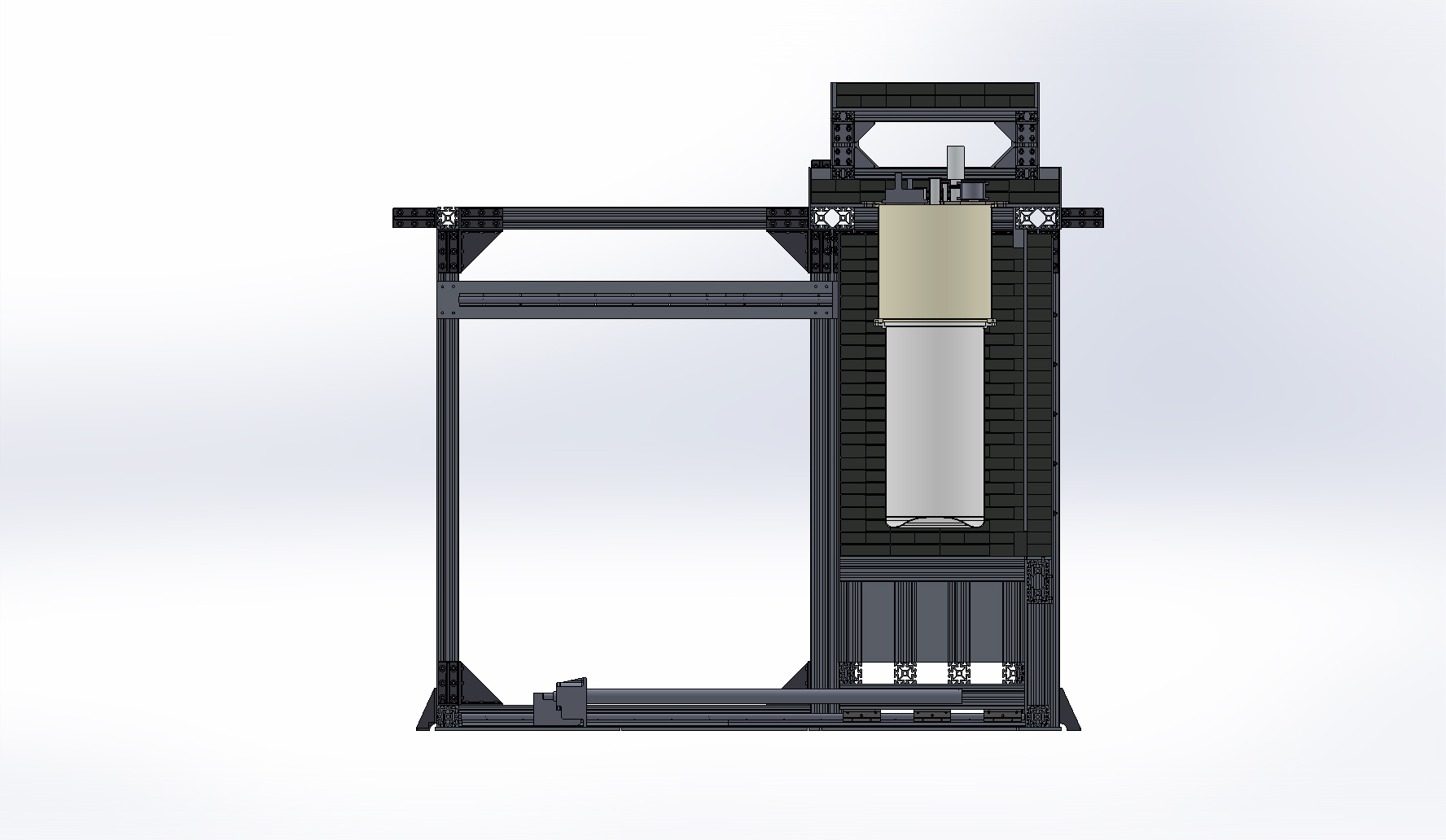}
    \caption{Draft renderings of the lead shield around the dilution refrigerator in open (left) and closed with cutaway (right) configurations.} 
    \label{fig:cad_fridge}
\end{figure}

Figure~\ref{fig:cad_fridge} shows a CAD rendering of the shield design. The shield is separated into bottom and top sections to accommodate the section of frame from which the dilution refrigerator hangs. The top is further separated into two sections, an upper roof and a lower skirt, with the gap between allowing for the fridge pulse tube and vacuum connections as well as experiment cabling. The upper sections create ``shadow shielding'' so that any straight ray drawn from the mixing chamber volume to the large gaps in the shield intersect with the top portion. The shield bottom consists of a single fixed wall attached to the dilution refrigerator frame, with the other four sides mounted to a linear motion system that allows access to the vacuum and IR shield cans and interior space and enables rapid shielded vs.\ unshielded comparison measurements. The lead is stepped where the two parts of the bottom shield meet to prevent line-of-sight gaps. 

Figure~\ref{fig:specvsshieldthick} also shows the estimated rate from radioactivity inside the dilution refrigerator already reported (assuming that the most significant sources, i.e. nearby PCBs and BeCu coaxial connectors, have been removed) and an estimate of the contribution from the shield itself, namely the aluminum support structure and \iso{Pb}{210} in the lead bricks. As Table~\ref{tab:ratevsshieldthick} shows, the expected contribution from \iso{Pb}{210} is greater than the expected residual environmental gamma flux, assuming a level of \SI{200}{Bq/kg}.

\subsection{Expected performance of superconducting qubits in the LBCF}
\label{sec:devices_in_lbcf}
The combination of overburden and lead shielding described in this work should reduce the total ionizing radiation interaction rate for devices inside the dilution refrigerator by approximately 95\% compared to an unshielded system operating at sea level.
We assume some care is taken to avoid introducing items with high radioactivity into the fridge such as ceramic PCBs. The primary purpose of this design is to enable the study of superconducting devices in a controlled radiation environment, 
in particular, testing of superconducting qubits and quasiparticle sensitive detectors for the effects of ionizing radiation. The simplest use case of the design includes A/B comparison testing of device performance between an above-ground, unshielded laboratory and the LBCF. PNNL has a similar model dilution refrigerator available on campus, above ground for such studies.

McEwen et.\ al.\ report time-correlated events of energy-relaxation errors for multiple qubits simultaneously across a qubit array, occurring at an average rate of $\lambda\sim1/(\SI{10}{s})$~\cite{mcewen2022}.  If the reported multi-qubit correlated error rate is directly proportional to the average ionizing radiation interaction rate, such a device in our shielded underground system would observe a burst rate of $\lambda\sim1/(\SI{200}{s})$. This estimate assumes that the radiation environment in their laboratory is similar to ours when unshielded and that there is essentially zero threshold energy to produce correlated error bursts. If instead there is a strong energy dependence, the rate reduction could be much less significant, as the reduction in cosmic-ray muons in the LBCF is modest and gammas cannot efficiently inject high energy into small pieces of silicon. For example, if the effective threshold for multi-qubit correlated error bursts is \SI{1}{MeV}, the reduction in rate would be roughly a factor of 4 only. These two contrasting scenarios create an opportunity to directly investigate details of the underlying mechanisms leading to ionizing radiation sensitivity for specific device designs.


We note the title of the McEwen et.\ al.\ report suggestively attributes the observed correlated error bursts to cosmic rays, though the authors do make clear that gamma rays from naturally occurring radioactivity in the environment can also contribute. Determining the relative influence of cosmic-ray secondary interactions and interactions from gamma-ray radiation is one of the goals enabled by the LBCF. By comparing measurements taken with the gamma shield open vs.\ closed, the contributions from gammas and from cosmic rays plus internal radioactivity can be cleanly separated. 

Harrington et.\ al.\ determined multi-qubit correlated error event rates from cosmic rays (at surface) and other sources separately by correlating a 5x5x0.35~mm$^3$ qubit device with a muon tracker~\cite{harrington2024}. They measure a total error event rate of $\sim$1/(100 s) and estimate the cosmic-ray-induced contribution is $\sim$1/(600 s) for their \SI{20}{mg}, vertically-oriented chip. 
Our model suggests a cosmic-ray threshold of $\sim$\SI{200}{keV}, with significant uncertainty 
given that we have simulated a device substrate with different dimensions (i.e., 2.5x5x0.38~mm$^3$) which will have a strong effect on cosmic-ray interaction rates. The total reported rate suggests a lower energy-threshold ($\lesssim$\si{50}{keV}) for gamma-rays, which is determined based on the mass of the qubit chip substrate and not the specific geometry. If the reported event rate in the Harrington et.~al.\ device is dominated by external gamma rays (rather than some internal source, such as a PCB), the expected event rate in the LBCF would be $\sim$1/hour and dominated by cosmic-ray-induced events.

\section{Discussion and conclusions}

\subsection{Further reduction in ionizing radiation}
\label{sec:further_reduction}
In this section we examine what steps would be required to further reduce ionizing radiation relative to the LBCF. For each successive reduction step, we target and mitigate the highest remaining source. Eventually we reach the background limits achieved by advanced fundamental physics detectors searching for neutrinoless double-beta decay events or interactions with galactic-halo dark matter particles. 

\paragraph{Cosmic-ray muons}
Cosmic-ray muons are the dominant source of ionizing radiation in devices operating in the underground shielded system. Further reduction would  require a site that is located deeper underground with greater overburden. A further reduction factor of 1000, at which point the residual muon-induced event rate would be a few percent of the rate from the fridge itself, would require a depth of roughly  \SI{1}{kilometer} water equivalent~\cite{laubenstein2004}. This depth is achievable in some of the shallower deep underground laboratories such as KURF (1450 m.w.e.)~\cite{kalousis2014} and WIPP (1585 m.w.e.)~\cite{esch2005}.

Depending on device and use case, an alternative to operating at a deeper site may be to employ a muon veto system as is commonly done for sensitive radiation detectors~\cite{knoll2010}. A muon veto would detect whenever a muon passed through some ``shell'' surrounding the device of interest, allowing for the rejection of any data generated during that time, or, alternatively, better characterizing device response to cosmic-ray interactions~\cite{harrington2024, li2024}. 
However, a veto system would not prevent errors in a superconducting qubit. While a veto system could provide a trigger to apply quantum fault mitigation~\cite{orrell2021}, for very long computational duration, such methods may still prove ineffective. This is another open research question that is intended for study at the LBCF.

\paragraph{Improved low background shielding}
After cosmic rays, three sources contribute roughly equally: residual external gammas, gammas from \iso{Pb}{210} in the shield lead~\cite{Keillor2017185}, and radioactivity inside the dilution refrigerator. 
Further reducing the ambient gamma flux is not as straightforward as increasing the shield thickness. As Table~\ref{tab:ratevsshieldthick} shows, the effect of the open top of the shield that allows free access to the vacuum flange is a significant source of the residual flux. Closing all of those openings would require a substantially more complex shield design.  Eventually, small gaps between lead bricks would become significant, requiring multiple layers to be staggered or redesigned with curved or ``chevron'' interfaces to remove any line-of-sight openings through the shield. 

Reducing the contribution from the shield itself is straightforward but costly. \iso{Pb}{210} is present at high levels in all recently-manufactured lead. There exist stockpiles of ``low background'' lead (often synonymous with ``Doe Run'' lead, the primary ore source) with \iso{Pb}{210} levels roughly an order of magnitude lower~\cite{Orrell20161271}. To further reduce this source, the innermost several centimeters would have to be replaced with so-called ancient lead, refined a sufficiently long time ago that most of the \iso{Pb}{210}, with 22~year half-life, has decayed away~\cite{knoll2010}. Because ancient lead is most often obtained from ancient shipwrecks, it is a very limited and correspondingly expensive commodity. To reach the lowest levels, an additional inner liner of several centimeters of copper is common.  Once the \iso{Pb}{210} is reduced by a factor of $\sim$10, the aluminum plates supporting the fridge and lead that are inside the shield become dominant.\footnote{Note that aluminum and steel plates and structural framing outside the shield are not included in our presented simulations as they contribute insignificantly to the total environmental gamma flux.} These would need to be removed from the design or, if not possible, replaced with a lower background material such as copper. 

\paragraph{Radiation sources inside the dilution refrigerator}
Further reducing the ionizing radiation event rate would require either an internal high density (e.g., tungsten or lead) shield, or modification of the instrument packaging and readout and the dilution refrigerator itself. From Table~\ref{tab:internalrate}, the most significant sources, assuming we have already removed ceramic interposers and BeCu coaxial connections on the device package, are:
\begin{itemize}
    \item Indium bump bonds. These are required for certain device designs (such as flip-chip connections). Should this become a hard limit, alternative designs not requiring bump bonds may be favored.
    \item Aluminum IR shields and vacuum can. These could be replaced with copper straightforwardly.
    \item Aluminum  and mumetal experiment shields. Here aluminum is chosen as a superconductor, so copper would not be a suitable replacement. A study would be required to evaluate the tradeoffs between residual radiation rate vs.\ lack of a superconducting shield, or identifying a suitable superconducting replacement with lower radioactivity than aluminum. We estimate the mumetal to contribute roughly an order of magnitude less than aluminum. 
    \item Cryogenic filters. If lower-background alternatives cannot be identified, a tradeoff study to evaluate the device performance with the filters moved  further from the device could be beneficial.
    \item Isolators located at the MXC plate. The assumed contamination level is derived from a measurement that yielded upper limits only~\cite{cardani2023}. A more sensitive assay is required to determine the true contribution. 
    \item BeCu in feedthroughs at the MXC plate. Moving the package as far as possible from the plate would have some benefit, otherwise custom feedthroughs without BeCu would be required. The research physics community has developed low-voltage, low-radioactivity connectors~\cite{busch2018} and cables~\cite{arnquist2023a}. However, such connectors and cables are likely not suitable for the cryogenic and RF-signal applications required for the case studied in this report. Further R\&D on material and instrumentation could likely address this need.

    \item Copper instrument package. The dominant contributors are \iso{Pb}{210} (out of equilibrium with \iso{U}{238}) with direct line-of-sight to the device and cosmogenic activation, both present in commercial copper. Both sources would be reduced significantly by replacing the package with electroformed copper~\cite{hoppe2014}.
\end{itemize}
These items account for $\sim$90\% of the estimated internal radioactivity. Significant further reduction would require construction of the experiment setup (including the dilution refrigerator unit) with ultra low-background materials and techniques similar to a dark matter experiment. For example, the SuperCDMS experiment locates the cooling elements of the dilution fridge outside the ionizing radiation shields, connected to the experiment volume by long tails~\cite{agnese2017}. This separates the devices from uncontrolled sources of radioactivity, with the tradeoff of significantly increased complexity and reduction in effective cooling power. 

As a partial alternative to complete redesign, an internal radiation shield may reduce the ionizing radiation backgrounds from the dilution refrigerator. Based on a prior, unpublished study, we estimate that a \SI{5}{cm} thick tungsten shield placed around the qubit package, with slits to allow cable connections, would attenuate the environmental gamma flux by a factor of $\sim$50. A similar reduction factor is expected for the residual flux after the steps above have been implemented. The addition of the shield mass to the mixing chamber stage would increase cooldown times. We also did not consider the intrinsic radioactivity of the tungsten itself in this estimate.  A recent report evaluates the levels of achievable purity in tungsten~\cite{hakenmuller2022}. In the prior study we chose tungsten over lead due to its higher gamma attenuation coefficient per unit thickness than lead. An ancient lead inner shield would have similar performance while being less massive and easier to machine. 

\paragraph{Current state-of-the-art limits}
The next generation of dark matter detectors expect to attain sufficient shielding and background reduction efficiency that they will be limited by \iso{Si}{32} in the silicon substrate~\cite{PhysRevD.95.082002, ORRELL20189, aguilar-arevalo2022}. \iso{Si}{32} has been measured in silicon CCDs to be approximately \mbox{10--100 /kg/day}~\cite{Aguilar-Arevalo2015, Aguilar-Arevalo2021}, corresponding to a reduction factor of $\sim$$10^{-6}$ to $10^{-7}$ compared to our estimates for unshielded surface operation. The feasibility to produce isotopically pure \iso{Si}{28} for ultra-low background detectors has been explored~\cite{ORRELL20189}, and other substrates without any long-lived isotopes, such as sapphire, might achieve even lower rates.  Attaining these levels of backgrounds requires tens of millions of dollars for shielding and custom ultra low-background components. These experiments operate in dedicated underground laboratories at depths of several kilometers water equivalent to obtain the necessary reduction in cosmic-ray muon fluence~\cite{mei2006}.


\subsection{Conclusion }\label{sec:conclusion}
In this report we have estimated the rate of ionizing radiation interactions in superconducting qubit devices\footnote{Although we have simulated a silicon substrate in this work, the general conclusions should be largely independent of the substrate material, so long as the substrate itself has comparably low levels of intrinsic radioactive contaminants.} from environmental sources and from internal radioactivity in a typical dilution refrigerator. We conclude that the rate of high energy interactions is strongly dominated by materials with high levels of radioactivity and within direct line-of-sight to the devices such as interposers composed of ceramic laminates and coaxial connectors containing BeCu. Other internal sources are subdominant regardless of effective energy threshold to typical laboratory gamma-ray and cosmic-ray secondary fluxes, which contribute roughly equally. We have presented the design methodology for a gamma-ray shield for the Low Background Cryogenic Facility in PNNL's 30~meters-water-equivalent Shallow Underground Laboratory. The shield and overburden combined reduce the total ionizing radiation interaction rate by approximately 95\% compared to an unshielded dilution refrigerator on the surface at sea level. Assuming multi-qubit correlated error event rates scale linearly with the ionizing radiation interaction rate, we expect that a representative superconducting quantum processor (estimates based on a Google Sycamore device~\cite{mcewen2022}) operated in the LBCF would exhibit error burst rates less than $\sim$1/200s, and a qubit device on a smaller, \SI{20}{mg} substrate could experience error burst rates less than $\sim$1/hour~\cite{harrington2024}. Further reduction by a factor of $\sim$10 could be achieved with the same design operated at a deeper ($\sim$1~km.w.e) site. Even further reduction would require a substantially more complex shield design and replacing some elements of the dilution refrigerator such as the aluminum IR shields due to relatively high naturally-occurring quantities of trace radionuclides in these materials. The methods outlined in this work and the simulated hit efficiency tables can be applied to quickly produce rough estimates of the rate of ionizing radiation-induced interactions for arbitrary materials inside a dilution refrigerator, given some knowledge of the level of intrinsic radionuclide contamination in that material. 

We believe this design concept provides utility for mitigation of highly-correlated catastrophic error bursts ~\cite{mcewen2022, wilen2021, harrington2024} as well as suppression of quasiparticle poisoning that reduces individual qubit coherence times~\cite{vepsalainen2020,cardani2021}. Both of these effects are associated with ionizing radiation interactions in superconducting quantum devices. 
The analysis presented in this report suggests that modest-sized shields located in shallow underground facilities are sufficient for providing an ionizing-radiation-reduced environment for the advancement and study of another 10- to 100-fold increase beyond current state-of-the-art superconducting quantum device coherence times~\cite{Place2021, wang2022}, assuming all other sources of decoherence are reduced commensurately. Such a facility also enables more controlled study of the characteristics of correlated errors and of tertiary effects of ionizing radiation such as its effect on two-level systems.

\acknowledgments
The authors would like to thank Joe Formaggio, Doug Pinckney, Mollie Schwartz, and Michael Kelsey for helpful discussions. 

The authors acknowledge support from the Pacific Northwest National Laboratory (PNNL) Laboratory Directed Research and Development (LDRD) program. The LDRD effort was responsible for the modeling and simulation efforts which resulted in a previous report~(PNNL-31996)~\cite{pnnl-31996}, the results of which were adapted for this report. The LBCF dilution refrigerator was provided by a PNNL institutional investment.  This work was supported by grants from the US Department of Energy Office of Nuclear Physics, and Office of High Energy Physics Quantum Information Science Enabled Discovery (QuantISED) program. Pacific Northwest National Laboratory is operated by Battelle Memorial Institute under contract no. DE-AC05-76RL01830 for the US Department of Energy. This research was supported by an appointment to the Intelligence Community Postdoctoral Research Fellowship Program at Massachusetts Institute of Technology administered by Oak Ridge Institute for Science and Education (ORISE) through an interagency agreement between the U.S. DOE and the Office of the Director of National Intelligence (ODNI). Work at MIT Lincoln Laboratory is supported under Air Force Contract No.\ FA8702-15-D-0001. Any opinions, findings, conclusions or recommendations expressed in this material are those of the authors and do not necessarily reflect the views of the U.S.\ Government.




\appendix

\section{Variance Reduction Techniques}\label{sec:append-variance}

Monte Carlo methods are, inherently, computationally and time intensive; so much so that obtaining statistically relevant results with singular compute nodes within days to weeks is challenging. This complication has been encountered in other cosmic background simulations; thus, highlighting the need to accelerate the simulation process. Variance reduction is the most common method of mitigating this problem.

Two methods of variance reduction were explored: one proposed by Battistoni in \cite{battistoni_monte_1997} which makes use of repeated geometries to increase the simulated count rate, and source biasing. Figure~\ref{fig:underground} depicts the model configuration of a dilution refrigerator housing sensitive instruments within a concrete shell and 19-m soil overburden.

The muon source term relative to a small object can be treated as an anisotropic plane source. As noted by Battistoni and Bielajew, repeated structures at a constant altitude are effectively equivalent because of the translation invariance of the problem. No impact on the angular or energy dependence of the incident particle counts was observed using this method.

The second method, source biasing, makes use of a maximum radial acceptance criterion whereby the direction of primaries at birth, $\hat{u}$, must be towards the recording volume. Using the center point of the tally volume, $v_t$, initial starting point of the primary, $v_p$, and a radial point perpendicular to the vector between the center of the tally volume and initial primary position ${\hat{u}_t}=v_t-v_p$ at a user specified distance, the maximum possible angular difference between the direction of the primary and ${\hat{u}_t}$ is computed as $\theta_a$ along with the angle between $\hat{u}$ and ${\hat{u}_t}$ as $\theta_u$. Only particles with $\theta_h \leq \theta_a$ are emitted and tracked. Unlike Battistoni's method, this technique was expected to reduce the number of low energy events. This bias was verified by applying the variance reduction technique to a simulated PVT muon counter, which was used in an earlier iteration of this study. Although the PVT panel itself is no longer used in this analysis, the derived geometric acceptance factor was carried forward. 

\begin{figure}[ht!]
    \centering
    \includegraphics[width=0.6\textwidth]{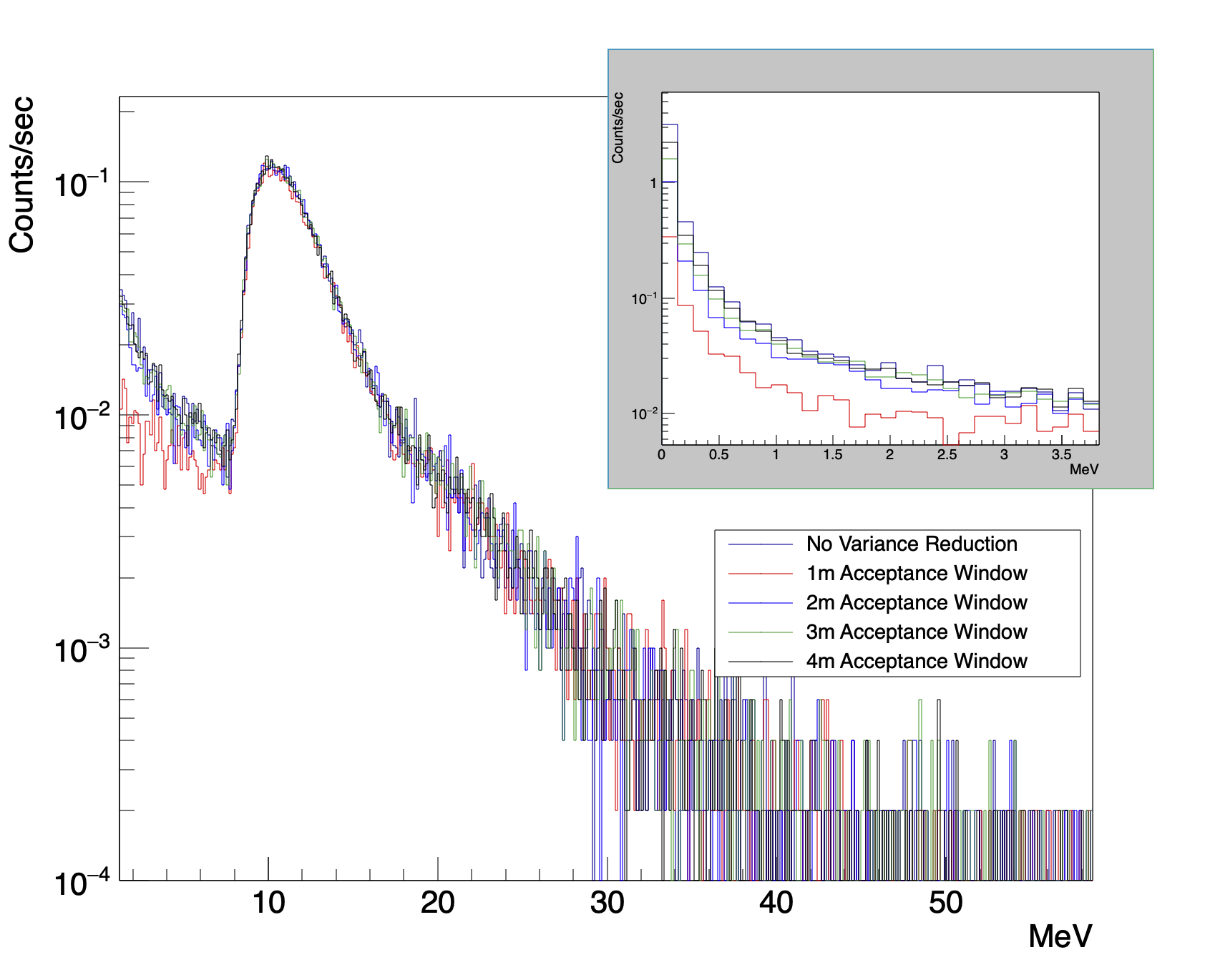}
    \caption{Comparison of the simulated muon spectrum interacting with a PVT scintillator panel as a function of the acceptance radius cut employed for variance reduction. A 1-meter cut shows significant divergence from the uncut spectrum at low energies, indicating that such a cut would introduce significant bias in the result. A 3-m cut was used for this analysis.}
    \label{fig:VR_radii}
\end{figure}

The radial acceptance range was evaluated at \mbox{1, 2, 3, 4 and $\infty$~meters}. A ``good'' variance reduction technique produces results identical to the full, unbiased simulation within statistical uncertainties but with lower variance for the same processing time. The comparison of results is provided in Figure~\ref{fig:VR_radii}. Virtually no impact to the muon peak or high-energy tail was observed, but a significant reduction in low-energy events (~80\% reduction in the lowest energy bin) was observed using a 1-m acceptance criteria applied to a 76-cm square PVT panel, as expected. This discrepancy was quickly reduced, however, by expanding the radial acceptance window; the 4-m window achieved ~80\% of the value observed with an open acceptance window in the lowest energy bin.
An improvement in events processed per CPU-hours of a factor of $\approx$170 was observed with a 1-m acceptance window and reducing inversely to the square of radius. 
We applied a radial acceptance cut of \SI{3}{m} to all simulations of cosmic rays in the SUL for this analysis. The simulated energy deposition spectrum in the unbiased simulation was equal to that obtained from the 3-m cut simulation within statistical fluctuations. 

\section{Procedure for fitting HPGe measurements}
\label{sec:hpgefits}
To determine the relative weights of the environmental sources in the simulation, we compared to the measurements taken with the HPGe detector in the SUL. We determined the individual nuclide concentrations that minimized the chi-squared of the fit of the simulated HPGe response to the collected data. The fit was performed using CERN ROOT's Minuit2 solver \cite{Hatlo:2005cj,Brun:1997pa} with 11 free parameters: an offset (pedestal) and linear scale factor to convert from energy to arbitrary analog to digital converter (ADC) counts, constant and square-root proportional terms for the energy resolution ($\sigma_E^2 = \sigma_0^2 + \sigma_1^2 E$), and linear scale factors for each of the seven isotopes listed above. 
Accurate simulation of a HPGe detector requires the optimization of several parameters. These parameters consist of values supplied by the vendor on detector design, consideration of detector aging features, and incomplete charge collection in portions of the crystal. The optimization of the dead layer thickness on the Ge crystal is particularly important, which strongly affects the HPGe response at low photon energies.

\begin{table}
    \centering
    \caption{Optimized and vendor provided parameters for Mirion GC14022 HPGe detector used in this work. The associated data is plotted in \protect{Figure~\ref{fig:totalgammaflux}}. }
    \label{tab:hpge_paramaters}
    \sisetup{table-format=3.3(1)}
    \begin{tabular}{l|ll}
        \toprule
        HPGe configuration & {Vendor Specified} & {Optimized} \\
        \midrule
        Ge crystal & &  \\
        \quad Diameter & - & 84\unit{mm} \\
        \quad Length & - & 84\unit{mm}\\
        \quad Outer dead layer thickness & 0.5 \unit{mm}& 1.2 \unit{mm}\\
        \quad Inner dead layer thickness & 0.3 \unit{\mu m} & 0.6 \unit{\mu m}\\
        \hline
        Crystal holder (copper) & & \\
        \quad Thickness & - & 7.5 \unit{mm} \\
        \hline
        End cap (aluminum) &  &  \\
        \quad Thickness &  -  &  0.5\unit{mm}  \\
        \quad Diameter &  108\unit{mm} &  108\unit{mm}  \\
        \quad Length & - &  159\unit{mm}  \\
        \quad Ge front to endcap distance & - & 7.5 \unit{mm}\\
         \hline
         Performance (at 1.33 MeV) & 140\% & 136\% \\
         \bottomrule
    \end{tabular}
\end{table}


The final optimized HPGe parameters used for this simulation are shown in Table~\ref{tab:hpge_paramaters} and the final resultant simulated spectrum along with the measured HPGe spectrum is shown in Figure~\ref{fig:totalgammaflux}. The data presented in Figure~\ref{fig:totalgammaflux} were normalized using the 63.9 live hours collection time (i.e., 99.4\% live during the data collection period) and results are reported in counts/second for each 0.36-keV-wide energy bin. The simulated fluxes recorded for each isotope are scaled by the corresponding fit amplitude and combined to produce a single reference for the total gamma flux in the SUL, which is then used as the input to estimate the gamma contribution to devices in the dilution refrigerator and evaluate shielding efficiency. 

\section{Variation of ambient gamma flux in different laboratories}
\label{sec:compareNaI}
\begin{figure}
    \centering
    \includegraphics[width=0.8\textwidth]{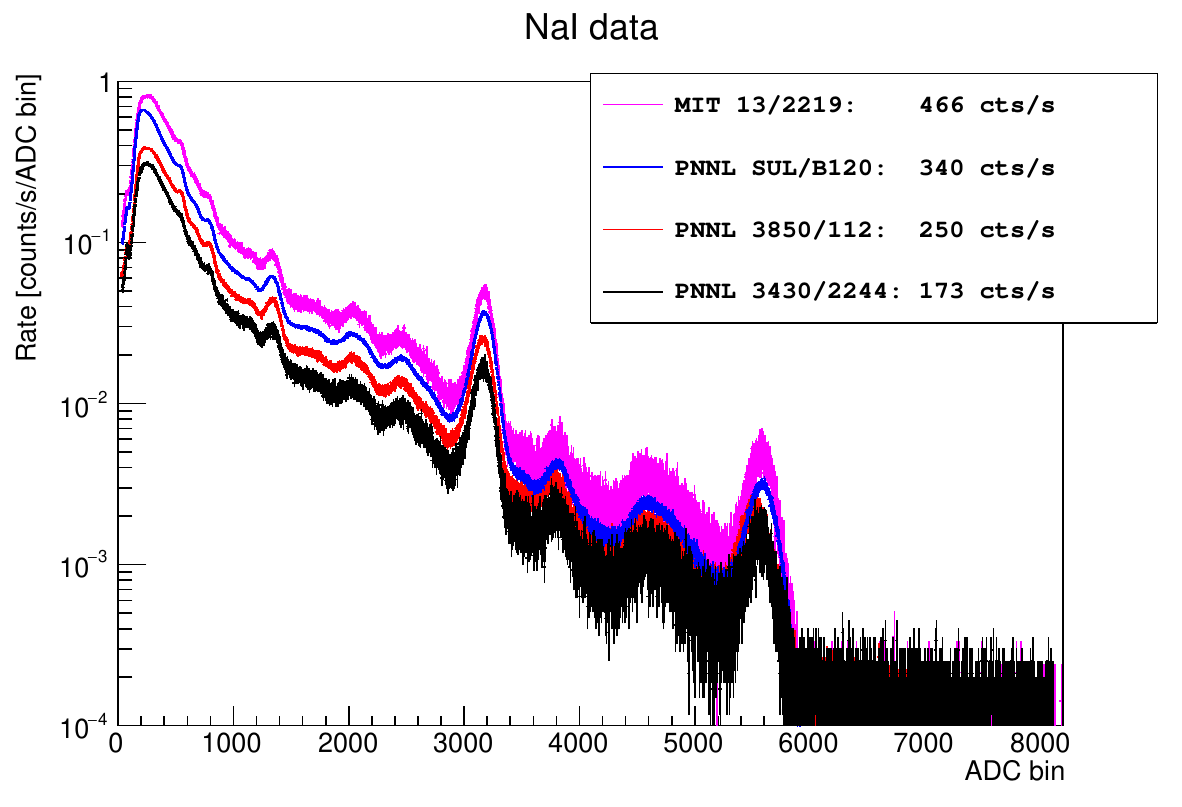}
    \caption{Comparison of background gamma spectra measured with a \SI{3}{inch} NaI detector at various laboratories, demonstrating the fairly small variation between different sites. The gains (horitontal axis) were adjusted manually to align the prominent \SI{1.4}{MeV} \iso{K}{40} peak. }
    \label{fig:compareNaI}
\end{figure}

In this work, we conclude that the radiation budget for superconducting devices is dominated by cosmic-ray secondaries and gammas from radioactivity in the surrounding laboratory. This conclusion is based on a single measurement of gammas in the SUL at PNNL. Figure~\ref{fig:compareNaI} shows measurements of ambient backgrounds in several locations using an NaI detector: one laboratory at MIT and three at PNNL, two at the surface and one in the SUL. The total interaction rate for each location is presented in the legend. Of the labs in the survey, the highest measured rate (at MIT) was 2.7 times larger than the smallest measured rate (in one of the PNNL surface laboratories). 

\section{Radiation transport hit efficiencies}\label{sec:append-hiteff}

We anticipate there are some researchers in the superconducting device community who do not have ready-built radiation transport Monte Carlo simulation models of their dilution refrigerator systems. However, they may nonetheless wish to answer specific questions of the following nature: \textit{If some device component might be radioactive at an estimated (or measured) level, then what is the potential impact on the superconducting device?} A full model and simulation is required to answer the question in an absolute sense. However, the \emph{relative} impact between different materials, at different locations, and of different radioactive background content, can be determined \emph{if and only if} a set of self-consistent radiation transport ``hit efficiencies'' is available. A hit efficiency is defined as the probability of radiation emitted from a given location and source in secular equilibrium to strike the device of interest. In other words, the contribution to the total ionizing radiation event (or dose) rate $R$ for a given component is given by
\begin{equation*}
    R = M \sum_{\mathrm{all}\ i} A_{i} E_{i,L}
\end{equation*}
where $M$ is the component's mass, $A_{i}$ is the specific activity of contaminant $i$ (e.g., \iso{U}{238}, \iso{Th}{232}) in the component, $L$ is the location of the component, and $E_{i,L}$ is the hit efficiency for contaminant $i$ at location $L$. 

Table~\ref{tab:hiteff} provides the hit efficiencies derived for locations throughout the dilution refrigerator. The ``Activation'' column in Table~\ref{tab:hiteff} refers to cosmogenic activation of copper at sea level as presented in Table~\ref{tab:radiopurity_lvls}. See the table notes for the relative activities of the isotopes considered. An example of how one may use Table~\ref{tab:hiteff} is as follows. To estimate the contribution of an alumina interposer board, we  take the radioactive assay values from Table~\ref{tab:radiopurity_lvls} to get:

\vspace{10pt}
\begin{tabular}{S@{\,kg$\quad\times\quad$}S[table-format=1.3]@{\,Bq/kg$\quad\times\quad$}S[table-format=1.1]@{\,/g$\quad=\quad$}S[table-format=1.5]@{\, cts/s/g\quad}c}
0.00078 & 5 & 7.3 & 0.028 & \iso{U}{238} \\
0.00078 & 0.066 &  5.2 &0.00027 & \iso{Th}{232} \\
0.00078 & 0.6 & 1.5 & 0.00070 & \iso{K}{40} \\
\hline
 \multicolumn{3}{r}{Total} & 0.029 & \\
\end{tabular}
\vspace{10pt}

\noindent which matches the value reported in Table~\ref{tab:internalrate}. 

As a second example, we can evaluate the contribution to the total background rate from 10 SMA connectors located at the CP stage to compare with the same number of connectors at the device package and at the MXC stage presented in Table~\ref{tab:radiopurity_lvls}. 

\vspace{10pt}
\begin{tabular}{S@{\,kg$\quad\times\quad$}S[table-format=2.1]@{\,Bq/kg$\quad\times\quad$}S[table-format=1.1e-1]@{\,/g$\quad=\quad$}S[table-format=1.2E-1]@{\, cts/s/g\quad}c}
0.023 & 23 & 1.7E-5 & 9.0E-6 & \iso{U}{238} \\
0.023 & 1.8 & 2.3E-5 & 0.95E-6 & \iso{Th}{232} \\
\hline
 \multicolumn{3}{r}{Total} & 1.0E-5 & \\
\end{tabular}

\vspace{10pt}
\noindent One can see a factor of 10 less than the same components located at the MXC plate, which we could infer directly by the ratio of the relevant entries in Table~\ref{tab:hiteff}, showing the utility of Table~\ref{tab:hiteff} for making quick, order-of-magnitude relative assessments.

\begin{table}
    \centering
    \caption{For a radioactive isotope (and its progeny in secular equilibrium in the case of \iso{U}{238} and \iso{Th}{232}) at a given location, average number ``hits'' depositing greater than \SI{3}{eV} in the silicon substrates per equilibrium decay (top) and average energy deposited per decay (bottom). Equivalently, conversion factors from decay rate to hit rate or dose. Values derived from Monte Carlo simulation. }
    \label{tab:hiteff}
    \begin{tabular}{l|rrrrrrr}
        \toprule
        Source location & \iso{U}{238} & \iso{Th}{232} & \iso{K}{40} & \iso{Co}{60} & \iso{Cs}{137} & \iso{Pb}{210} & Activation \\        
        \midrule
        & \multicolumn{7}{c}{Hit efficiency, \si{1/g/s/Bq}} \\
        \hline
        Bump bonds & 8.3E+2	& 6.6E+2 & 5.4E+1 & 5.6E+1& 6.4E+1 & \multicolumn{2}{c}{\iso{In}{115}: 5.7E+1} \\
        Interposer board & 7.3E+0 & 5.2E+0 & 1.5E+0 & 3.1E-1 & 8.3E-1 & 1.5E+0 & 4.2E-1 \\ 
        Package & 7.3E-2 & 6.0E-2 & 1.2E-2 & 2.1E-2 & 9.8E-3 & 8.0E-3 & 1.4E-2 \\ 
        Package Connector Inside & 8.4E-1 & 5.2E-1 & 1.8E-1 & 5.3E-2 & 7.5E-2 & ~ & ~ \\ 
        Package Connector Outside & 1.4E-2 & 1.7E-2 & 9.4E-4 & 1.4E-2 & 4.8E-3 & ~ & ~ \\ 
        Experiment stage & 7.3E-4 & 1.0E-3 & 4.5E-5 & 9.1E-4 & 2.3E-4 & 2.5E-6 & 5.2E-4 \\ 
        Experiment shield & 2.2E-4 & 2.8E-4 & 1.3E-5 & 2.5E-4 & 8.1E-5 & 0.0E+0 & 1.5E-4 \\ 
        Mixing Chamber Stage & 1.2E-4 & 1.6E-4 & 8.8E-6 & 1.5E-4 & 4.4E-5 & 1.8E-7 & 8.7E-5 \\ 
        Cold Plate Stage & 1.7E-5 & 2.3E-5 & 1.1E-6 & 2.3E-5 & 6.8E-6 & 1.4E-8 & 1.3E-5 \\ 
        Still Stage & 7.3E-6 & 9.3E-6 & 5.8E-7 & 9.5E-6 & 2.6E-6 & 4.8E-9 & 5.4E-6 \\ 
        4K Stage & 1.6E-6 & 2.3E-6 & 1.3E-7 & 2.7E-6 & 4.1E-7 & 0.0E+0 & 1.5E-6 \\ 
        50K Stage & 4.6E-7 & 7.4E-7 & 2.1E-8 & 8.2E-7 & 1.9E-7 & 3.1E-9 & 4.4E-7 \\ 
        Vacuum Flange & 2.6E-7 & 3.3E-7 & 1.5E-8 & 4.0E-7 & 8.6E-8 & 0.0E+0 & 2.3E-7 \\ 
        Still Can & 6.0E-5 & 8.1E-5 & 4.3E-6 & 7.4E-5 & 2.1E-5 & 7.5E-8 & 4.4E-5 \\ 
        4K Can & 3.0E-5 & 3.9E-5 & 2.1E-6 & 3.6E-5 & 1.1E-5 & 9.7E-9 & 2.1E-5 \\ 
        Lower 50K Can & 2.5E-5 & 3.1E-5 & 1.8E-6 & 2.9E-5 & 9.1E-6 & 9.7E-9 & 1.7E-5 \\ 
        Upper 50K Can & 9.3E-7 & 1.3E-6 & 3.6E-8 & 1.5E-6 & 4.4E-7 & 0.0E+0 & 7.9E-7 \\ 
        Lower Vacuum Can & 1.7E-5 & 2.3E-5 & 1.4E-6 & 2.1E-5 & 7.6E-6 & 0.0E+0 & 1.2E-5 \\ 
        Upper Vacuum Can & 6.3E-7 & 1.0E-6 & 8.7E-8 & 1.1E-6 & 2.1E-7 & 0.0E+0 & 5.7E-7 \\ 
        \midrule
        & \multicolumn{7}{c}{Dose efficiency, \si{keV/g/s/Bq}} \\
        \hline
        Bump bonds & 1.9E+6 & 1.6E+6 & 1.3E+4 & 4.0E+3 & 8.9E+3 & \multicolumn{2}{c}{\iso{In}{115}: 6.0E+3} \\
        Interposer board & 2.7E+3 & 2.3E+3 & 3.3E+2 & 3.7E+1 & 1.3E+2 & 4.2E+2 & 2.8E+1 \\ 
        Packge Inner Surface & 2.3E+4 & 1.9E+4 & 1.9E+2 & 3.6E+1 & 7.6E+1 & 1.7E+3 & 2.2E+1 \\ 
        Package & 2.0E+1 & 1.8E+1 & 2.6E+0 & 3.6E+0 & 1.3E+0 & 2.7E+0 & 2.0E+0 \\ 
        Package Connector Inside & 3.0E+2 & 2.3E+2 & 3.7E+1 & 8.4E+0 & 1.1E+1 & ~ & ~ \\ 
        Experiment stage & 1.0E-1 & 1.4E-1 & 7.8E-3 & 1.5E-1 & 2.5E-2 & 1.5E-4 & 7.9E-2 \\ 
        Experiment shield & 2.9E-2 & 3.8E-2 & 1.5E-3 & 4.2E-2 & 9.6E-3 & 0.0E+0 & 2.3E-2 \\ 
        Mixing Chamber Stage & 1.6E-2 & 2.1E-2 & 1.4E-3 & 2.4E-2 & 4.7E-3 & 9.8E-6 & 1.3E-2 \\ 
        Cold Plate Stage & 2.2E-3 & 2.9E-3 & 1.9E-4 & 3.3E-3 & 6.4E-4 & 7.5E-7 & 1.8E-3 \\ 
        Still Stage & 9.6E-4 & 1.2E-3 & 9.7E-5 & 1.4E-3 & 2.4E-4 & 1.4E-7 & 7.2E-4 \\ 
        4K Stage & 2.1E-4 & 3.2E-4 & 2.3E-5 & 3.7E-4 & 4.0E-5 & 0.0E+0 & 1.9E-4 \\ 
        50K Stage & 6.8E-5 & 1.0E-4 & 4.7E-6 & 1.2E-4 & 1.4E-5 & 1.3E-8 & 5.8E-5 \\ 
        Vacuum Flange & 3.0E-5 & 5.8E-5 & 2.1E-6 & 6.2E-5 & 9.6E-6 & 0.0E+0 & 3.2E-5 \\ 
        Still Can & 7.9E-3 & 1.0E-2 & 6.6E-4 & 1.2E-2 & 2.4E-3 & 3.3E-6 & 6.3E-3 \\ 
        4K Can & 3.8E-3 & 5.2E-3 & 3.5E-4 & 5.6E-3 & 1.3E-3 & 4.2E-7 & 3.0E-3 \\ 
        Lower 50K Can & 3.0E-3 & 4.3E-3 & 3.1E-4 & 4.7E-3 & 9.5E-4 & 2.2E-7 & 2.5E-3 \\ 
        Upper 50K Can & 1.2E-4 & 1.7E-4 & 8.0E-6 & 2.3E-4 & 4.3E-5 & 0.0E+0 & 1.2E-4 \\ 
        Lower Vacuum Can & 2.1E-3 & 3.1E-3 & 2.4E-4 & 3.2E-3 & 8.7E-4 & 0.0E+0 & 1.7E-3 \\ 
        Upper Vacuum Can & 8.7E-5 & 1.4E-4 & 1.2E-5 & 1.7E-4 & 2.0E-5 & 0.0E+0 & 8.2E-5 \\ 
        \bottomrule
    \end{tabular}
\end{table}

\section{Transmon qubit device fabrication and assay methodology}
\subsection{Fabrication}
\label{sec:append-qubitfab}
In fabrication of the assayed transmon qubit device, Al base metallization was grown via thermal evaporation from an effusion cell in an ultra-high-vacuum (UHV) molecular-beam epitaxy (MBE) system on a high-purity, single-crystal silicon wafer. Embedding circuitry (feedlines, readout resonators, bias lines, etc.) was patterned using optical lithography and wet-etched with Aluminum Etchant Type A (Transene Company, Inc.). Josephson junctions (JJs) were fabricated using the Dolan-bridge technique, in which a suspended bridge produces shadows from two angled-evaporation steps~\cite{dolan1977}. In the MIT LL process, this is achieved with a mask stack consisting of ZEP520A resist (ZEONREX Electronic Chemicals) on a thin layer of Ge, supported by a sacrificial layer of electron-beam resist copolymer MMA(8.5)/MAA EL9 (MicroChem). The ZEP520A is exposed by a Vistec EBPG-5200 electron beam pattern generator and developed, after which the pattern is transferred to the Ge via a $\mathrm{CF}_4$ reactive-ion etch. The MMA/MAA layer is then cleared in the vicinity of the pattern by an $\mathrm{O}_2$ plasma. An \emph{in situ} Ar ion-milling was performed before angled deposition of the first Al electrode of the Josephson junctions. Thermal oxidation in a dedicated chamber grew the junction barrier prior to a final angled deposition of Al to create the second junction electrode. Finally, airbridge crossovers to link ground planes in coplanar waveguide structures were added using optical lithography and electron-beam evaporation of Al. The wafers were then coated with protective organic photoresist for dicing into 2.5x5~mm chips. The chips are finally cleaned with ACS-grade solvents and mounted into packages for testing.



\subsection{Assay}
\label{sec:append-qubitassay}
Measurements were conducted using an Agilent 8900 triple quadrupole inductively coupled plasma mass spectrometer (QQQ-ICP-MS; Agilent Technologies, Santa Clara, CA, USA) with integrated autosampler, Pt skimmer and sampler cones, s-lens, and standard electron multiplier detector. A quartz double-pass spray chamber and a 100~$\mu$L$\cdot$min$^{-1}$ microflow perfluoroalkoxy alkane (PFA) nebulizer (Elemental Scientific, Omaha, NE, USA) were used as the sample introduction system. Optimal tuning parameters for signal sensitivity and stability were determined using a 0.1~ng$\cdot$g$^{-1}$ $^{205}$Tl standard.


Sample preparation and analysis involved dissolving three weighed qubits individually in a mixture of Optima grade nitric and hydrofluoric acids in the presence of a known amount of non-natural $^{229}$Th and $^{233}$U tracer. Once dissolved, samples were dried down and resuspended in 2\% nitric acid before being introduced into the ICP-MS. Measurements of each sample were conducted as a triplicate scan, and the instrumental error was derived from one standard deviation of the three scans. Sample quantitation was conducted using standard isotope dilution methods described elsewhere~\cite{WOS:000348040900014,WOS:000596707400007}, and detection limits were determined as 3~times the standard deviation of the procedural blanks that were carried through the same process as the samples. All chemistry was conducted in a Class~10 laminar flow hood using acid leached and validated PFA vials to prevent contamination during the assay. All reagents were made with 18.2~M$\Omega$$\cdot$cm deionized water.






\clearpage
\bibliographystyle{unsrtnattrunc} 
\bibliography{references}

\end{document}